\documentclass[natbib]{svjour3}                     %
\smartqed
\usepackage{graphicx}
\usepackage{amsmath}
\usepackage{amssymb}
\usepackage{epstopdf}
 \journalname{Space Sci Rev}
\newcommand{\aap}{{Astron. Astrophys.}}
\newcommand{\apj}{{Astrophys. J.}}
\newcommand{\apjl}{{Astrophys. J.}}
\newcommand{\apjs}{{Astrophys. J. Supp. Series}}

\newcommand{\mnras}{{MNRAS}}
\newcommand{\nature}{{Nature}}
\newcommand{\nat}{{Nature}}

\newcommand{\prd}{{Phys. Rev. D}}
\usepackage{txfonts}
\DeclareSymbolFont{cmletters}{OML}{cmm}{m}{it}
\DeclareMathSymbol{v}{\mathalpha}{cmletters}{"76}
\begin{document}

\title{Radio Pulsars
}

\author{V.S. Beskin         \and
        S.V. Chernov\and
        C.R.~Gwinn\and
         A. Tchekhovskoy%
}

\institute{V.S. Beskin \at
                Lebedev Physical Institute, Russian Academy of Sciences\\
                Leninskii prosp. 53
                Moscow, 119991 Russia\\
\emph{and:}\\
                 Moscow Institute of Physics and Technology \\
                 Institutskii per. 9,
                 Dolgoprudnyi, Moscow Region, 141700 Russia \\
              \email{beskin@td.lpi.ru}           %
           \and
           S.V. Chernov \at
              Lebedev Physical Institute, Russian Academy of Sciences\\
               Profsoyuznaya st. 84/32, Moscow, 117997 Russia\\
              \email{chernov@td.lpi.ru}           %
           \and
           C.R. Gwinn \at
              University of California, Santa Barbara\\
              Physics Dept., Broida Hall\\
              Santa Barbara, California, 93106 United States of America\\
              \email{cgwinn@physics.ucsb.edu}           %
           \and
        A. Tchekhovskoy \at
              University of California\\
              Department of Astronomy\\
              Berkeley, California, 94720, United States of America \\
              \email{   atchekho@berkeley.edu}           %
}

\date{Received: date / Accepted: date}

\maketitle

\begin{abstract}
Almost 50 years after radio pulsars were discovered in 1967, our understanding
of these objects remains incomplete. On the one hand, within a few years
it became clear that
neutron star rotation gives rise to 
the extremely stable sequence of radio pulses, that the
kinetic energy of rotation provides the reservoir of energy, and that electromagnetic fields are the braking mechanism.
On the other hand, 
no consensus regarding the mechanism of coherent radio emission or the conversion of electromagnetic energy to particle energy yet exists.
In this review, we report on three aspects of pulsar structure that have seen recent progress:
the self-consistent theory of the magnetosphere of an oblique magnetic rotator;
the location, geometry, and optics of radio emission;
and evolution of the angle between spin and magnetic axes.
These allow us to take the next
step in understanding the physical nature of the pulsar activity.
\end{abstract}

\clearpage

\section{Introduction}\label{sec:intro}

Radio pulsars are the archetypal observed neutron stars. Their discovery
at the end of the 1960s ~\citep{nature1968} was definitely one of the major astrophysical events of the 20th
century. Their discovery confirmed the theoretical prediction of neutron stars in the 1930s~\citep{Landau, BaZwi}. 
Neutron stars have mass $M$ of about $1.2$--$2.0\ M_{\odot}$, near the
Chandrasekhar mass limit $1.4\ M_{\odot}$; and radius $R$ of only $10$--$15$ km. They result
from the collapse of typical massive stars in the final stage of their evolution~\citep{ShapTeuk}; or from white
dwarfs, when accretion from a companion
star pushes them over the Chandrasekhar limit~\citep{Whe73,Bai90,Nom91,Schwab}. These formation mechanisms provide the simplest explanation
for both the observed short spin periods $P$ to as small as $P = 1.39$ ms, and superstrong
magnetic fields with \mbox{$B_0 \sim 10^{12}$ G}.

Most radio pulsars are solitary. Of the more than 2400 pulsars known by
the end of 2014, only about 230 were members of binary systems.\footnote{See ATNF catalog:
http://www.atnf.csiro.au/people/pulsar/psrcat/}  Even in binary systems, mass transfer from the companion star to the
neutron star is negligible. The radio luminosities of pulsars are low relative to the sensitivities of even the largest radiotelescopes,
so that our catalog of pulsars is not complete even to a distance of a kpc. 
Because the Milky Way is an order of magnitude larger, we can observe only
a small fraction of ``active'' pulsars. 
Because the duration of the active life of pulsars is small,
the total number of extinguished pulsars in our Galaxy must
be about \mbox{$10^{8}$--$10^{9}$} \citep{ManTable}.

\section{Theoretical Overview}

\subsection{Early Pulsar Paradigm --- Vacuum Dipole}\label{sec:early}

The basic physical processes determining the observed activity of radio pulsars were
understood almost immediately after their discovery~\citep{Pacini, Gold}. In particular, it quickly became clear
that the highly-regular pulsed radio emission that gives rise to their name is related to the rotation of
neutron stars. Furthermore, it was evident that radio pulsars are powered by the rotational energy of the neutron
star, and the mechanism of energy release is related to their superstrong magnetic fields,
with $B_0 \sim 10 ^{12}$ G. The Larmor
formula for energy loss form a magnetic dipole provides an estimate of energy losses\ \citep{LLField}:
\begin{eqnarray}
W_{\rm tot} &=& -I_{\rm r}\Omega\dot\Omega \label{eq:rotenergyloss} \\
&\approx& \frac{1}{6}\frac{B_0^2\Omega^4R^6}{c^3}\sin^2\chi 
\label{eq:wmd}
\end{eqnarray}
where $I_{\rm r} \sim MR^2$ is the moment of inertia of the neutron star, $\chi$
is the angle between the magnetic dipole axis and the spin axis, and  $\Omega = 2\pi/P$
is the angular velocity of neutron star rotation. Finally, the strength of the magnetic 
field at the polar cap is $B_0$. 

For most pulsars, energy losses
range from  $10^{31}$--$10^{34}\ {\rm erg\ s}^{-1}$ and can reach $10^{38}$--$10^{39}\ {\rm erg\ s}^{-1}$ for
very young, fast pulsars, such as the Crab and Vela pulsars. These energy losses correspond to
the observed spin-down rate ${\rm d}P/{\rm d}t \sim 10^{-15}$, or to the spin-down time
$\tau_{\rm D} = P/2\dot P \sim$ $1$--$10\ {\rm Myr}$.

After the measurement of the rotational slow-down $\dot P$ of the Crab pulsar \citep{RichCom69},
it was quickly realized that:
\begin{description}
\item{$\bullet$} the rate of the energy loss of the rotating neutron star 
$W_{\rm tot} \approx 5 \times 10^{38}\ {\rm erg\ s}^{-1}$ \eqref{eq:rotenergyloss}
coincides with the power required to illuminate the 
Crab Nebula \citep{Gold69}, and
\item{$\bullet$} 
the dynamical age of the radio pulsar
$\tau_{\rm D} = P/2\dot P \approx 1000$ years
coincides with the
explosion of the historical supernova AD 1054 that brought the Crab Nebula into existence \citep{Com69}.
\end{description}
These associations cemented the identification of pulsars as rotating neutron stars.
In contrast to these phenomena, radio emission amounts to only $10^{-4}$--$10^{-6}$
of total energy losses. For most pulsars this corresponds to $10^{26}$--$10^{28}\ {\rm erg\ s}^{-1}$, 
5--7 orders of magnitude less than the luminosity of the Sun.

\subsection{Electron-positron generation}\label{sec:epgeneration}

\citet{GJ} showed shortly after the discovery of pulsars that a pulsar's rotating magnetic 
field will acquire a corotating charge density that opposes induced electric fields and 
$\vec J\times \vec B$ forces. As \citet{stur71} quickly realized, individual photons can 
generate electron-positron pairs when they cross lines of the magnetic field, by the process 
$\gamma+B \rightarrow e^{+}+e^{-} + B$. The photon energy must exceed the threshold 
$2 m_{\rm e}c^2$. The probability per-unit-length for conversion of a photon with energy
${\cal E}_{\gamma}$ far above this threshold propagating at an angle of $\theta$ to the magnetic field ${\bf B}$ is~\citep{blp}
\begin{equation}
w = \frac{3 \sqrt{3}}{16 \sqrt{2}} \,
\frac{e^3 B\sin\theta}{\hbar m_{\rm e}c^3}
\exp\left(-\frac{8}{3}\frac{B_{\hbar}}{B\sin\theta}
\frac{m_{\rm e}c^2}{{\cal E}_{\gamma}}\right).
\label{d8}
\end{equation}
Here, the characteristic value $B_{\hbar} = m_{\rm e}^2c^3/e\hbar \approx 4.4 \times 10^{13}$ G 
is the magnetic field for which the energy gap between two Landau levels reaches the rest energy 
of an electron: $\hbar \omega_{B} = m_{\rm e}c^2$. As gamma-quanta are radiated by particles moving
along the curved magnetic field lines, one can evaluate the photon free path as~\citep{rs}
\begin{equation}
l_{\gamma} \approx \frac{8}{3 \Lambda} \, R_{\rm c} \, \frac{B_{\hbar}}{B} \, \frac{m_{\rm e}c^2}{{\cal E}_{\gamma}}.
\label{bh}
\end{equation}
Here $R_{\rm c}$ is the curvature radius and 
$\Lambda \approx 20$ is the logarithmic factor. As $l_{\gamma} \ll R$ for high enough 
photon energy, the vacuum magnetosphere of a neutron star with  magnetic field $B_0 \sim 10^{12}$ G  
is unstable to the generation of charged particles.

\begin{figure}
  \includegraphics[scale=0.39,clip=true,trim=2.5cm 5.1cm 2.5cm 5.3cm]{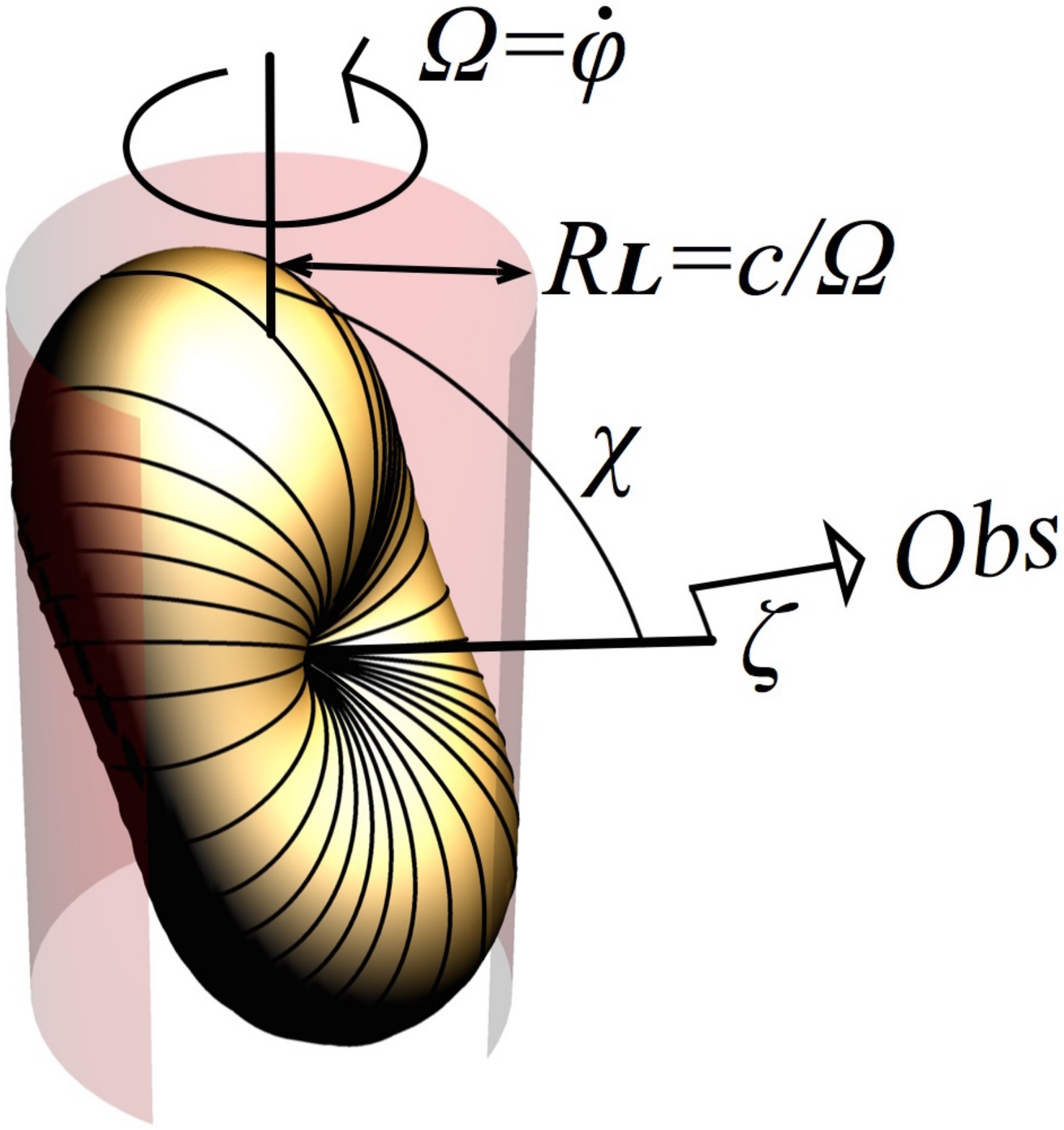}
  \includegraphics[scale=0.27,clip=true,trim=0.1cm 3.0cm 0.9cm 3.0cm]{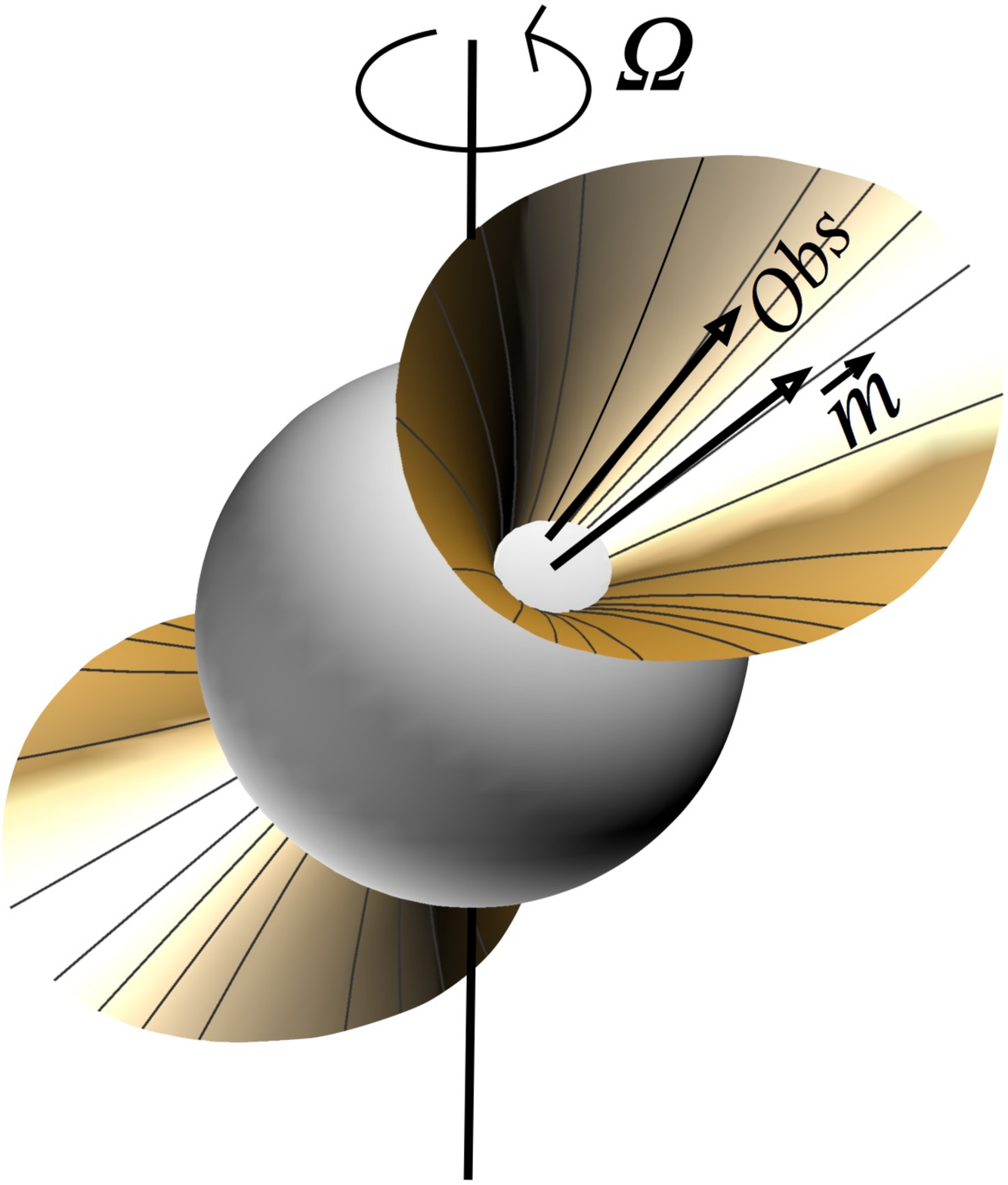}
\parbox{50mm}{\caption{Vacuum dipole model of a pulsar, showing the region of closed 
                       magnetic field lines and the light cylinder.
                       The angle between the rotation and magnetic poles is $\chi$, often called the ``inclination angle''; and 
                       the minimum angle between the line of sight and the magnetic axis is $\zeta$, sometimes called the ''impact angle''.
                       Sometimes $\chi$ is denoted by $\alpha$, and $\zeta$ by $\beta$. See \citet{Lyn88}.
\label{fig:vacuumdipole}}}
              \hfill\parbox{60mm}{\caption{Intersection of the cone of the last open field lines 
                                   with the surface of the neutron star, showing the polar cap.
                                   The vector $\vec m$ is the axis of the magnetic dipole field,
                                   and Obs points toward the observer.
\label{fig:polarcap}}}
\end{figure}

\begin{figure}

  \includegraphics[scale=.3]{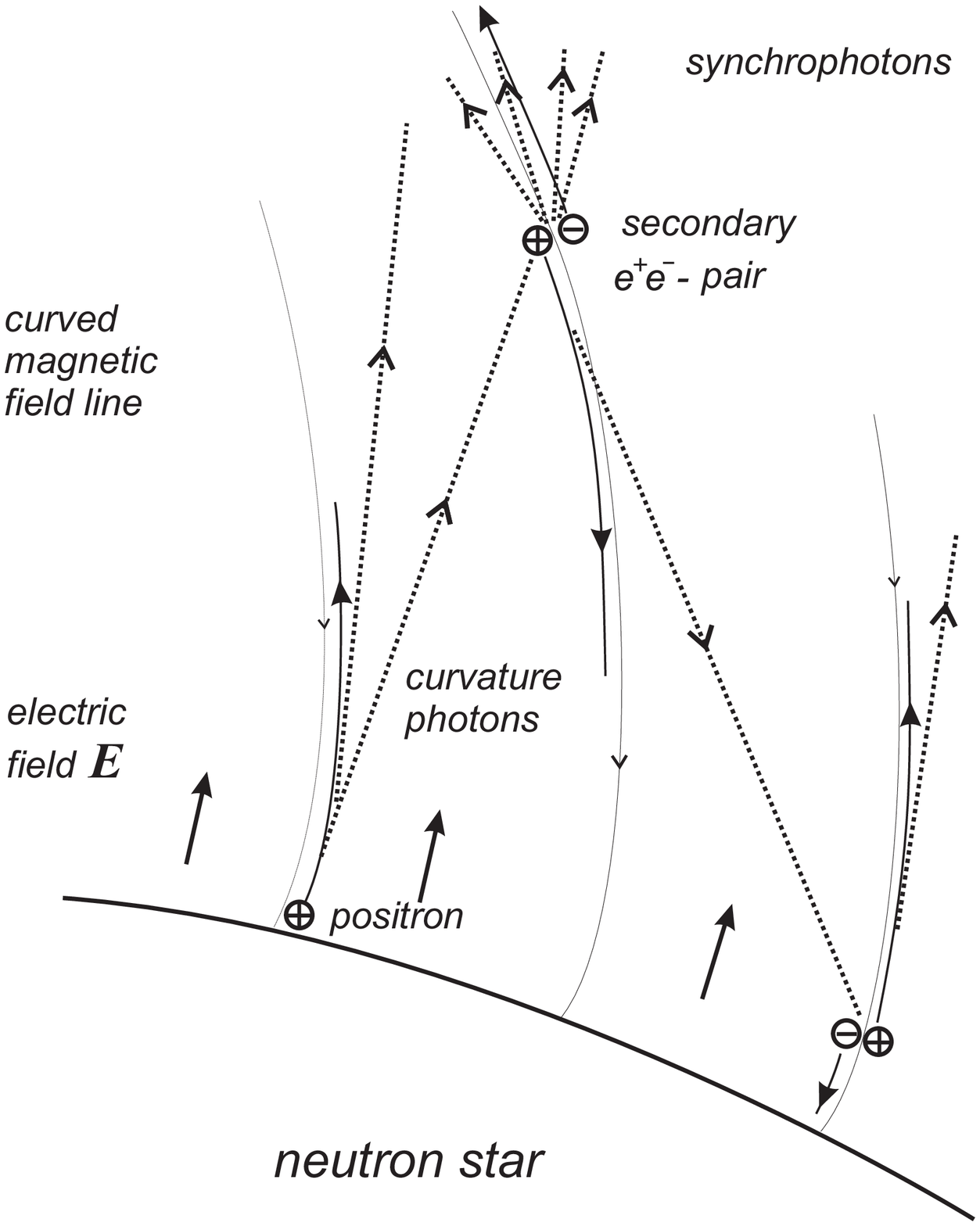}
   \includegraphics[scale=.35]{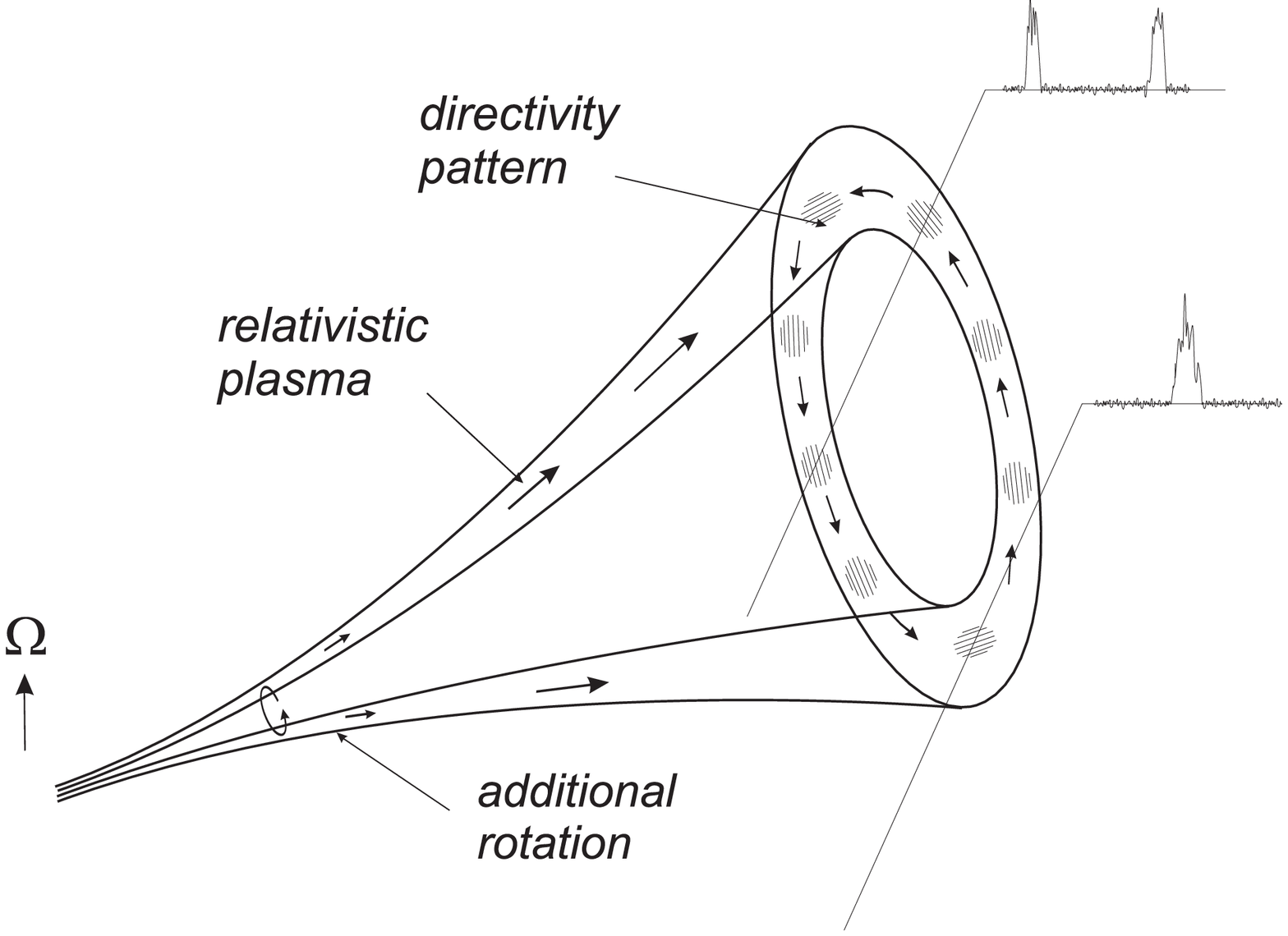}
\parbox{50mm}{\caption{Structure of the particle generation region. The primary
        particles are accelerated along the curved magnetic field lines and begin
        to radiate hard $\gamma$-rays. These curvature photons ({\it dotted lines})
        propagating in the curved
        magnetic field reach the particle generation threshold
        and create electron-positron pairs. Taken from~\citet{be1999}.
        }\label{fig:01}}
\hfill\parbox{60mm}
{\caption{The hollow cone model. If the intensity of the radio emission
                   is directly connected with the outflowing
                   plasma density, radio emission must decrease near the magnetic pole.
                    Consequently, we expect a double profile when the line of sight passes near the magnetic pole,
                   and a single profile when it passes further away. Taken from~\citet{be1999}.
                   }\label{fig:02}}       %
\end{figure}

In the very strong magnetic field of the neutron star, charged particles can move only along magnetic field lines.
Therefore 
two substantially different regions must develop in the pulsar magnetosphere: regions
of open and closed magnetic field lines (see Figure~\ref{fig:vacuumdipole}, \ref{fig:polarcap}). 
Closed magnetic field lines do not intersect the light cylinder, where co-rotation speed equals that of light,
at radius $R_{\rm L} = c/\Omega$ ($\sim 10^{10}$ cm for ordinary pulsars).
Particles on these field lines
turn out to be captured. Open field lines intersect the light cylinder,
and particles on these field lines 
can travel to infinity. Consequently, plasma must be continuously regenerated near the magnetic poles of a 
neutron star (see Figure~\ref{fig:01}).

In addition to the primary plasma generated by individual photons and the magnetic field, as discussed above,
a secondary plasma forms from the longitudinal electric field (which accelerates particles up to energies
high enough to radiate hard $\gamma$-quanta), as first indicated by~\citet{stur71} and then studied in more 
detail by~\citet{rs}, as well as by Eidman's group~\citep{ei}. The continuous escape of particles along the 
open field lines leads to formation of a strong electric field along the magnetic field. This longitudinal 
electric field forms in the vicinity of the magnetic poles. The secondary plasma-generation condition 
determines its height. Another model, based on the assumption of free particle escape from the neutron 
star surface,was first studied by Arons' group~\citep{arons1, arons2, arons3}, and recently in more detail 
by~\citet{Sobyanin1, Sobyanin2, Laietalcap, Tim, Beloborodov} and \citet{TimArons}.

\subsection{Hollow-cone model}\label{sec:HC}

The hollow-cone model~\citep{RadCoo69,Dyks04} explains the basic observed properties of 
radio emission in the context of the above particle generation processes, without reference 
to a microphysical model for that emission. This model, already proposed at the end of the
1960s, perfectly accounts for the basic geometric properties of the radio emission.
This model proposes that outflowing plasma launches radio emission tangent to open magnetic field lines
at a particular altitude above the surface of the neutron star.
The characteristic frequency of radiation may depend on altitude:
the ``radius-to-frequency mapping'' \citep{rs}.
Plasma density and geometry of open field lines define a ``directivity pattern''.
The observed average pulse is a cut across this directivity pattern.

Secondary particle generation is impossible in a nearly rectilinear
magnetic field because, first, little curvature radiation is emitted; and second,
photons emitted by relativistic particles propagate at small angles to the magnetic field.
Therefore, as shown in Figure~\ref{fig:02}, in the central region of the open magnetic field
lines, a decrease in secondary plasma density is expected.
If we make the rather reasonable assumption that radio emission is less when the
outflowing plasma density is less, the intensity of radio emission must decrease in the center 
of the region of open field lines, corresponding to the center 
of the directivity pattern. Therefore, if without going into details,\footnote{Actually, the mean
profiles have a rather complex structure, see e.g.,~\cite{rk83, rk90, lg-s}} we should expect a
single (one-hump) mean profile in pulsars in which the line of sight intersects the directivity
pattern far from its center and the double (two-hump) profile for the central passage. This is
precisely as observed in reality~\citep{lg-s}.

\section{Observational Overview}

Pulsars take their name from their remarkably stable periodic emission. The rotation frequency 
of the pulse train is the angular velocity of the neutron star. Folding the observed pulse train 
at this fundamental frequency yields an average pulse profile. In most cases this profile is 
extremely stable, both in form, and in arrival phase at the rotational frequency. This stability 
allows for precision timing of pulsars, with remarkable applications in structure and evolution 
of stellar systems containing pulsars, and in tests of special and general relativity~\citep{Cmz}. 
The stability of the mean profile
suggests that rotation carries the line of sight through a beam of emitted radiation locked to the surface of the neutron star,
and that relatively permanent features of the neutron star and its co-rotating magnetosphere determine the shape of that beam.

However, pulsar emission
shows a remarkable degree of variability on all timescales,
extending from nanoseconds to months or years.
The stable pulse profiles that characterize that stability appear only after 100 or more pulses are added together,
for pulsars strong enough to detect variability of emission.
Indeed, \citet{pop06a} suggest that individual micropulses are the ``atoms'' of pulsar emission.
For those who seek to understand emission processes of pulsars,
as well as those who merely wish to exploit pulse stability for other scientific goals,
pulse variability can provide crucial insights.

Observations of pulsars occupy a multi-dimensional space.
The fundamental observables include intensity and polarization of electromagnetic radiation,
as function of 
time in pulse phase and over many pulses.
These fundamental observables show both deterministic and random properties,
with random properties in particular showing variations over all timescales.
At radio wavelengths, pulsar spectra are nearly power-law, but comparisons
of pulse shape and structure among wavelength ranges yields great insight into emission geometry and processes \citep[see, for example,][]{Shea03,Lomm07,Har08,Stra13}.

Among the important quantities derived from observations of radio emission from pulsars
are the spindown rate, polarization as a function of pulse phase, size of emission region,
and evolution of angle between the spin and magnetic dipole.

\section{Magnetosphere of an Oblique Magnetic Rotator}\label{sec:magrot}

A pulsar represents an elegant problem in electrodynamics: 
a rotating, conducting sphere with a dipole magnetic field \citep{Bes13}.
This simple picture is complicated by the necessity of a corotating charge distribution,
the roles of open and closed field lines, 
and energy transport by an outflowing wind \citep{GJ}.
As we summarize in this section,
models have progressed from analytic studies of aligned rotators with simple magnetic field configuration and massless charges
to self-consistent models including oblique magnetic fields, realistic particle masses, and a range of length scales.

\subsection{Current Losses}\label{sec:paradigm}

If the pair creation process is sufficiently effective, magnetic dipole radiation will not carry energy away from the rotating neutron star,
because the plasma that fills the magnetosphere
fully screens any low-frequency radiation from the neutron star~\citep{bgi83, bgi93,mps99}. 
However, in this case, electric currents extract rotational energy from the neutron star, through the Amp\`ere force of one current on another.
The currents in question are those along magnetic field lines in the magnetosphere and across the pulsar's polar cap,
acting together with those responsible for the magnetic field of the neutron star.
Just as in the case of magnetic dipole radiation, energy release from the rotating neutron star is related
to the electromagnetic energy flux given by the Poynting vector, and the total energy losses
can be again estimated using the Larmor formula, \eqref{eq:wmd}.

\begin{figure}
\begin{minipage}[t]{0.49\linewidth} %
\begin{center}
 \includegraphics[scale=.32]{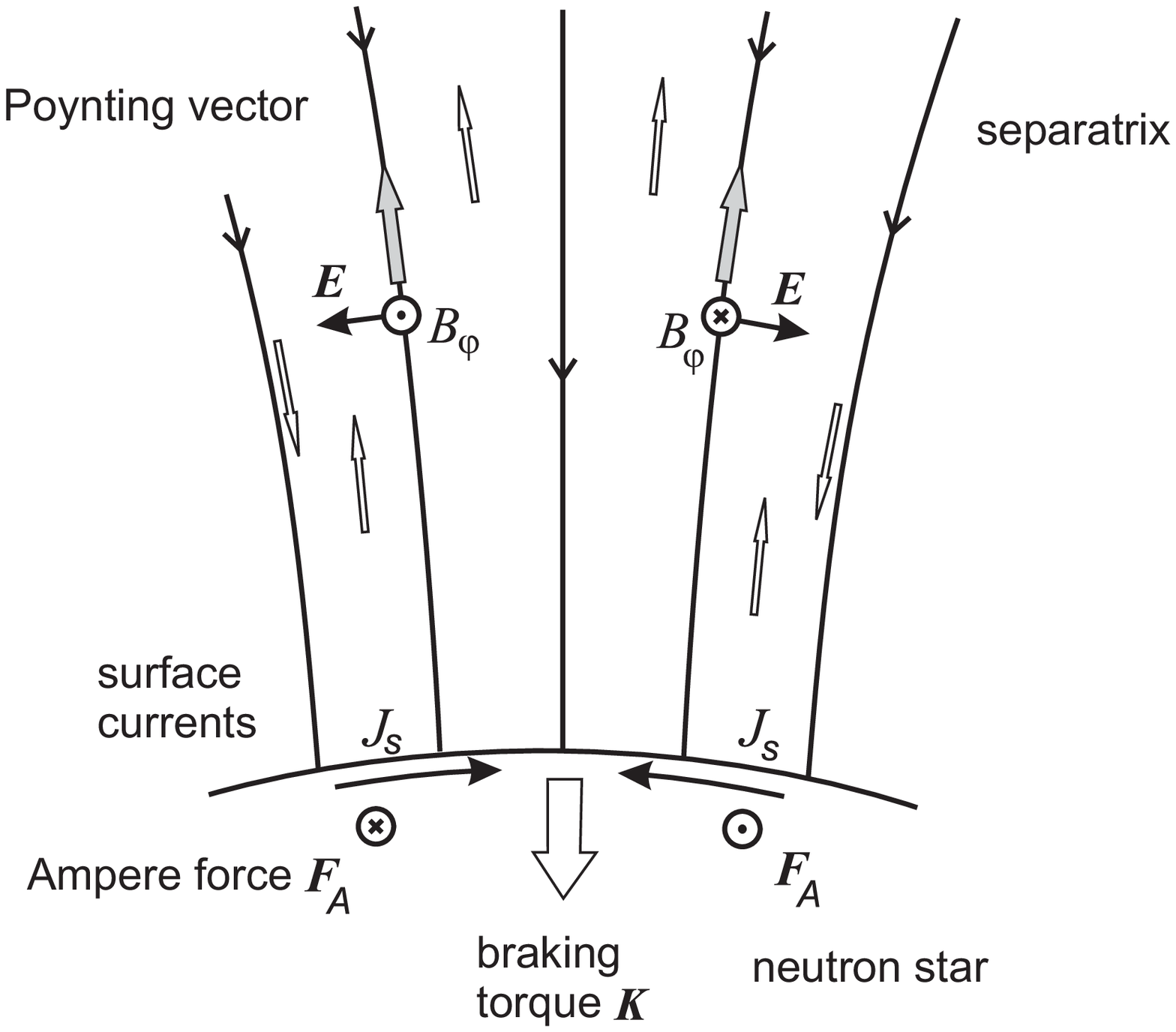}
{\caption{Schematic view of the axisymmetric polar cap showing magnetospheric current density (open arrows),
surface currents, Amp\'ere force on surface currents, and braking torque. Here only the symmetric current $i_{\rm s}$
is present. Taken from~\citet{be1999}.
        }\label{fig:2_11}}
\end{center}
\end{minipage}
\hspace{0.01\textwidth} %
\begin{minipage}[t]{0.49\linewidth} %
\begin{center}
 \includegraphics[scale=.45]{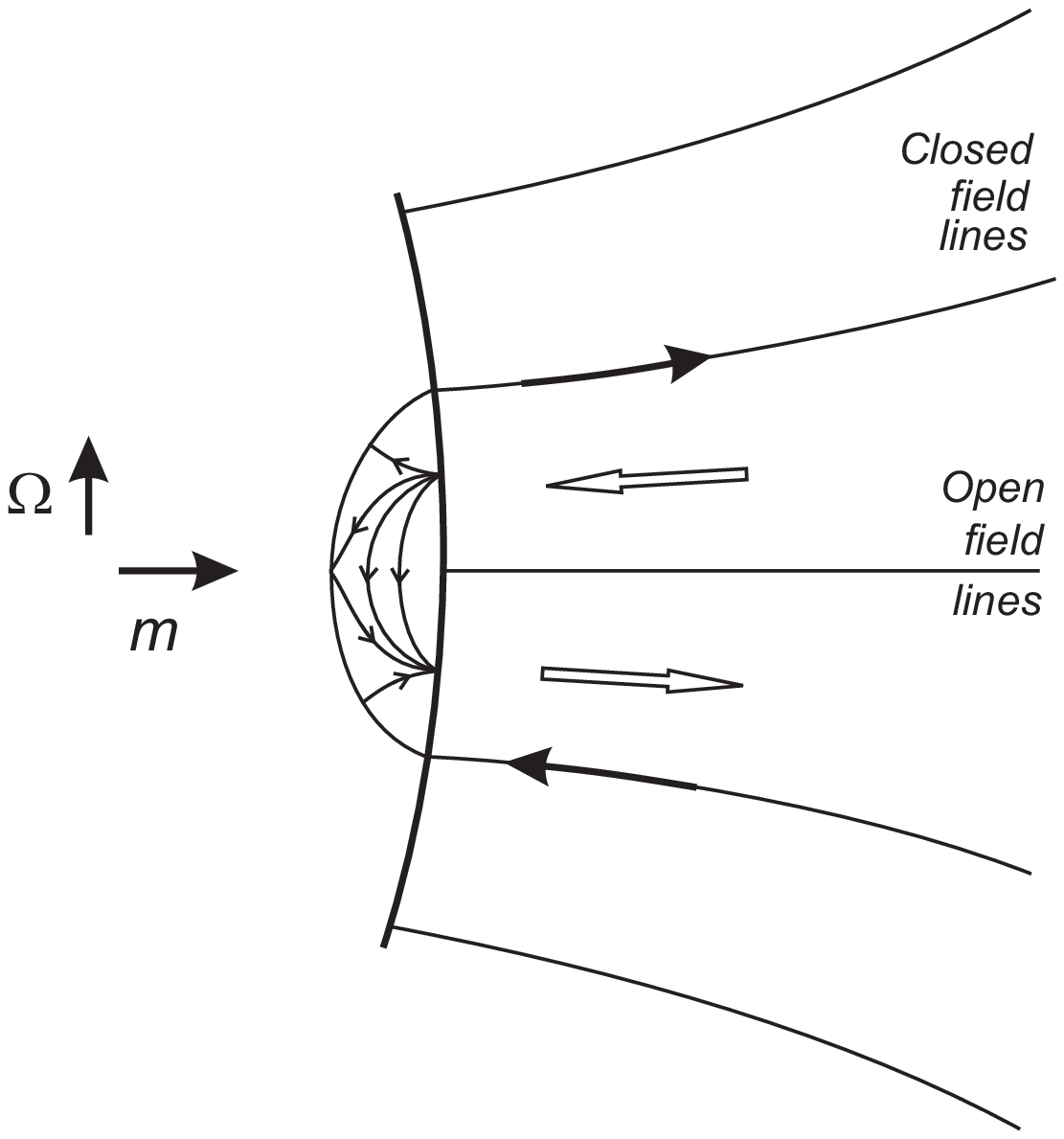}
{\caption{For the orthogonal rotator only antisymmetric current $i_{\rm a}$ (i.e., the
current having different direction in the north and south parts of the polar cap) takes place.
The structure of the surface currents within the polar cap and along the separatrix is also shown.
Taken from~\citet{Bes13b}.
}\label{fig:2_12}}       %
\end{center}
\end{minipage}
\end{figure}

The braking torque ${\bf K}$ of the Amp\`ere force
results in the following time evolution of the angular velocity $\Omega$ and the
inclination angle $\chi$:
\begin{eqnarray}
I_{\rm r} \, \dot{\Omega}
& = & K_{\parallel}\cos{\chi}+K_{\perp}\sin{\chi},
\label{18'} \\
I_{\rm r}\Omega \, {\dot\chi}
& = & K_{\perp}\cos{\chi}-K_{\parallel}\sin{\chi},
\label{19'}
\end{eqnarray}
where two components of the torque ${\bf K}$ parallel and perpendicular to the magnetic
dipole ${\bf m}$ can be written in the form~\citep{bgi93}
\begin{eqnarray}
K_{\parallel} & = &
- c_{\parallel} \frac{B_{0}^{2}\Omega^{3}R^{6}}{c^{3}} i_{\rm s},
\label{16'} \\
K_{\perp} & = & - c_{\perp} \frac{B_{0}^{2}\Omega^{3}R^{6}}{c^{3}}
\left(\frac{\Omega R}{c}\right)i_{\rm a}.
\label{17'}
\end{eqnarray}
Here the coefficients $c_{\parallel}$ and $c_{\perp}$ are factors of the order of unity
dependent on the profile of the longitudinal current and the form of the polar cap. 

The scalar current from the polar cap $i$ has been divided into 
symmetric and antisymmetric contributions, $i_{\rm s}$ and $i_{\rm a}$, depending upon 
whether the direction of the current is the same in the north and south parts of the polar 
cap, or opposite. For an axisymmetric rotating neutron star ($\chi=0$, Figure~\ref{fig:2_11}), 
we have $i_{\rm a} = 0$  and $i_{\rm s} = 1$  \citep{GJ}. Conversely, for the 
orthogonal rotator (Figure~\ref{fig:2_12}) we have $i_{\rm s} = 0$ and $i_{\rm a} = 1$.
Here we apply normalization to the Goldreich-Julian current, $I_{\rm GJ} = \pi R_{0}^2 j_{\rm GJ}$,
where $R_{0} \approx R (\Omega R/c)^{1/2}$ is the polar cap radius, and 
$j_{\rm GJ} = <|{\bf \Omega}\cdot{\bf B}|>/2\pi$ (with scalar product) is the mean current
density within the polar cap.
Note that for $i_{\rm s} \approx i_{\rm a} \approx 1$, 
\eqref{16'} and \eqref{17'} imply that:
\begin{eqnarray}
K_{\perp} \sim \left(\frac{\Omega R}{c}\right) K_{\parallel}.
\label{iais}
\end{eqnarray}
Therefore, $K_{\perp} \ll K_{\parallel}$. We will use these expressions in the 
following sections.  %

If we suppose that, in reality, the longitudinal current $j$ is
determined by the local charge density $\rho_{\rm GJ} = -{\bf \Omega}\cdot{\bf B}/2 \pi c$,
and note that $\rho_{\rm GJ}$ 
is proportional to $\cos\chi$ in the vicinity of the polar cap, one can write down
\begin{eqnarray}
i_{\rm s} & = & i_{\rm s}^{A} \cos\chi,
\label{16new} \\
i_{\rm a} & = & i_{\rm a}^{A} \sin\chi.
\label{17new}
\end{eqnarray}
Consequently, the relations \eqref{18'}--\eqref{19'} can be rewritten in the form
\begin{eqnarray}
I_{\rm r}\dot{\Omega}
& = & K_{\parallel}^{A} + [K_{\perp}^{A}-K_{\parallel}^{A}]\sin^2\chi,
\label{8'} \\
I_{\rm r}\Omega {\dot\chi}
& = & [K_{\perp}^{A}-K_{\parallel}^{A}]\sin\chi\cos\chi.
\label{9'}
\end{eqnarray}
As we see, both expressions contain the factor $[K_{\perp}^{A}-K_{\parallel}^{A}]$. This implies
that the sign of $\dot\chi$ is given by the $\chi$-dependence of
the energy losses \citep{Phi13}. In other words, the inclination angle $\chi$ will evolve to $90^{\circ}$ 
(to counter-alignment) if the total energy losses decrease for larger inclinational angles, and 
to co-alignment if they increase with inclination angle.

Because the plasma filling the pulsar magnetosphere is secondary (in other words, it is produced by the
primary particles accelerated by the longitudinal electric field), at any point out to
the light cylinder, the energy density of the electromagnetic field must be much larger
than the energy density of the magnetospheric plasma. For the same reason, energy transport is given by the Poynting vector
(see Figure~\ref{fig:2_11}).

\begin{figure}
  \includegraphics[scale=.4]{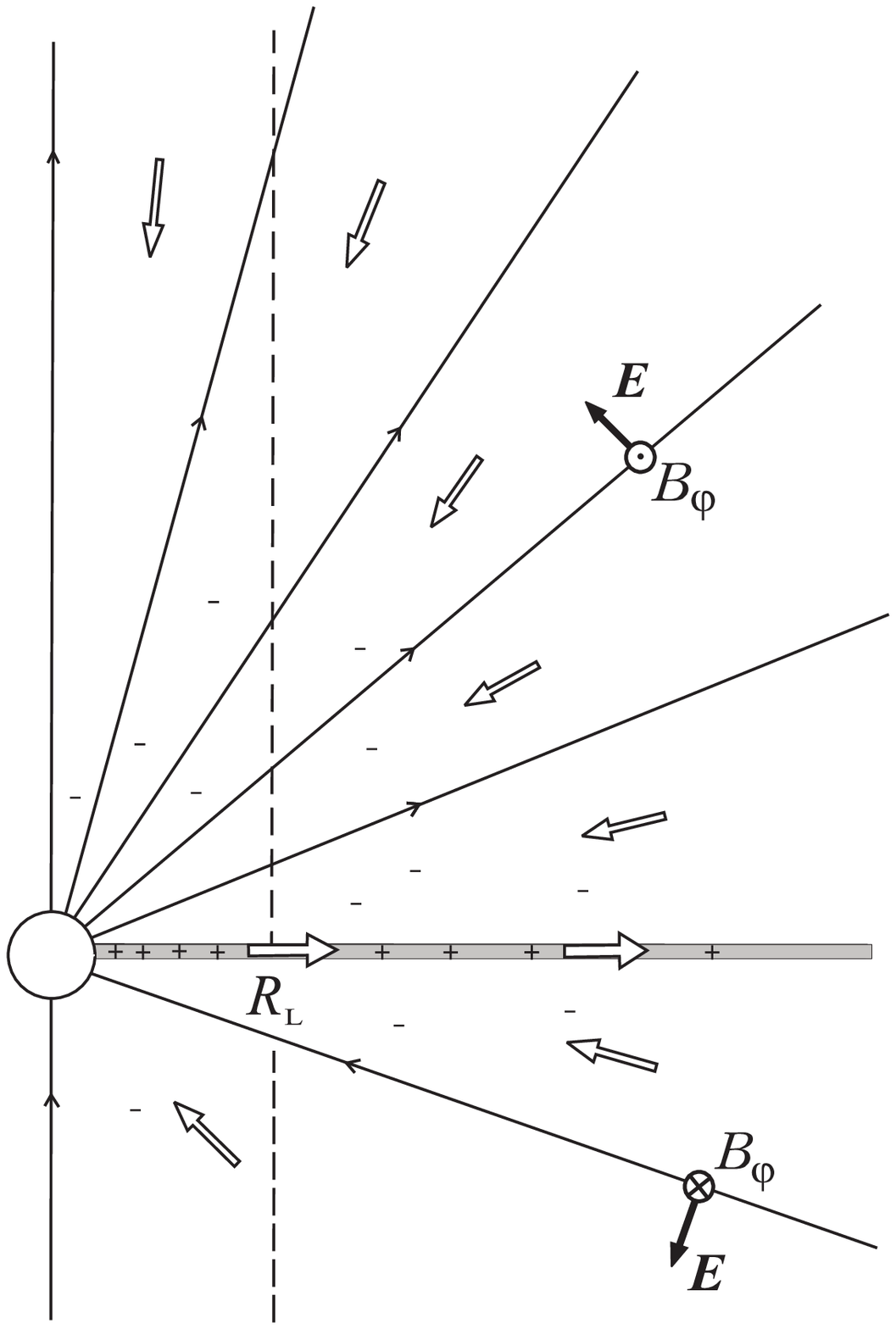}
  \includegraphics[scale=.4]{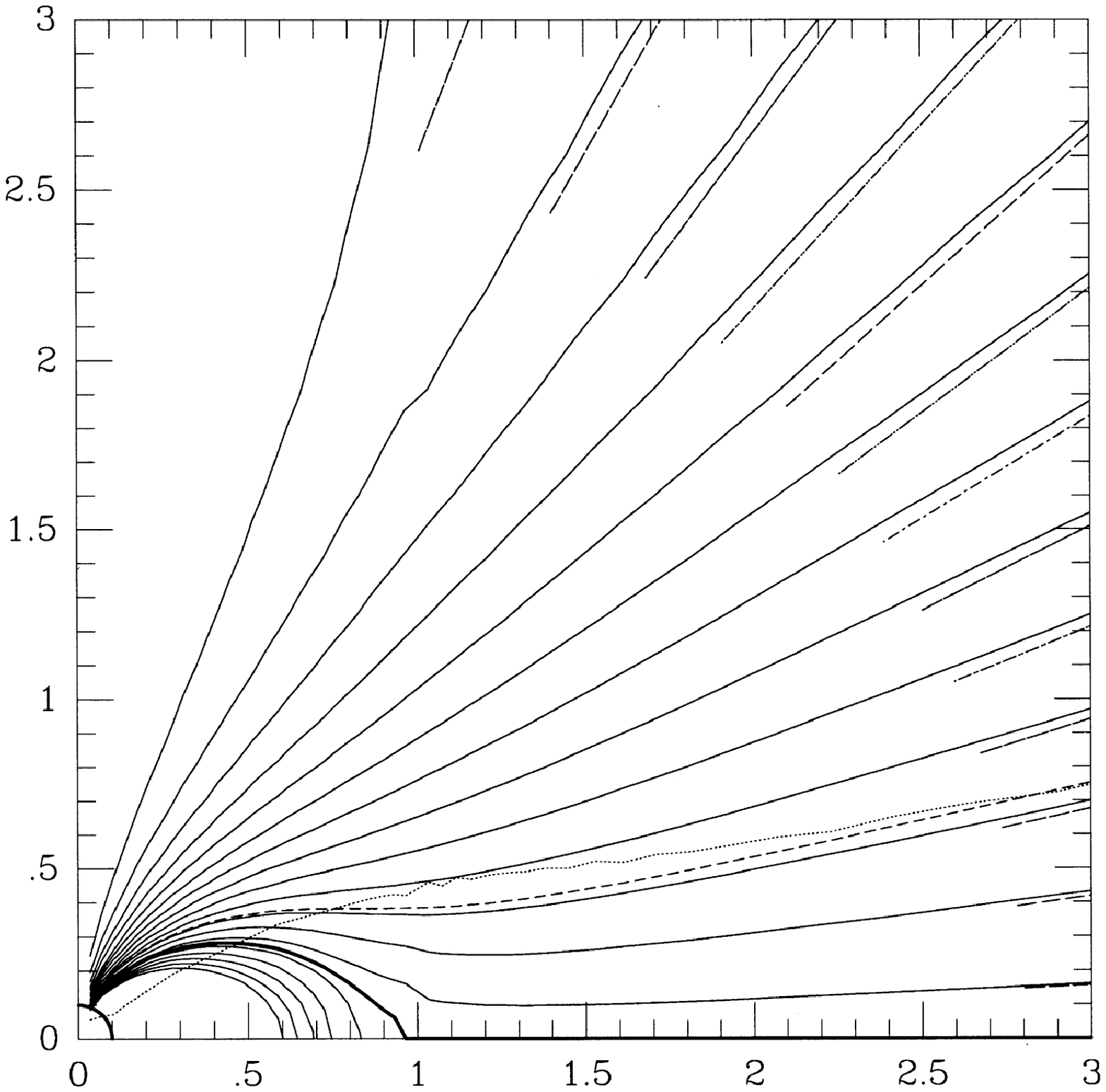} \\
\parbox{50mm}{\caption{The Michel split-monopole solution, in which electric
                   field $E_{\theta}$ has magnitude equal to the toroidal
                   magnetic field $B_{\varphi}$. This solution requires a
                   conducting current sheet, to close electric
                   currents ({outlined arrows}). Taken from \citet{be1999}.}\label{fig:michel_monopole}}
              \hfill\parbox{60mm}{\caption{Magnetospheric structure
         in the model of~\citet{ckf99}. The last open field line is assumed to coincide with
         the equator.}
\label{fig:04}}
\end{figure}

\subsection{Split-Monopole Model}\label{sec:split_monopole}

The remarkable analytical solution
found by~\citet{mich73b} serves to illustrate the transport of energy by Poynting flux. 
In the force-free approximation, when massless
charged particles move radially with the velocity of light, and with the Goldreich-Julian current density
$j_{\rm GJ} = \rho_{\rm GJ}c$, 
a split-monopole magnetic field 
is the exact solution to the Maxwell's equations, both inside and beyond the light cylinder
(see Figure~\ref{fig:michel_monopole}). 
The monopolar magnetic field is split so that  the magnetic flux converges in the southern hemisphere and diverges in the northern one.
In this solution, Amp\`ere forces from longitudinal currents along magnetic field lines, and
from corotation currents from rotating charge density, are fully compensated. 

In the Michel split-monopole solution, the electric field ${\bf E}$ has only a $\theta$-component,
and is equal in magnitude to the toroidal component of the magnetic field:
\begin{equation}
B_{\varphi} = E_{\theta}
= - B_0\left(\frac{\Omega R}{c}\right)\frac{R}{r}\sin\theta .
\label{eq:split_monopole_Etheta_Bphi}
\end{equation}
At distances larger than the light cylinder radius, this magnetic field becomes larger than  the poloidal
magnetic field $B_{\rm p} = B_0(R/r)^2$. On the other hand, in this solution the total magnetic
field remains larger than the electric field everywhere, so that the fields form electromagnetic waves only at infinity.

Because magnetic flux converges in the lower hemisphere and diverges in the upper one in the split-monopole solution,
a current sheet must lie in the equatorial plane
(see Figures~\ref{fig:michel_monopole}, \ref{fig:04}). 
This sheet closes the longitudinal electric currents elsewhere in the magnetosphere.
This structure of the magnetic field and current sheet has been confirmed
numerically~\citep{ckf99, ok03, gruz1,
k06, mck06, tim06}.

Using Eqn. (\ref{eq:split_monopole_Etheta_Bphi}), one easily finds that the Poynting vector
${\bf S} = (c/4\pi) {\bf E}\times{\bf B}$ is:
\begin{equation}
S(\theta)
= \frac{B_{0}^2 c}{4 \pi}\left(\frac{\Omega R}{c}\right)^2\frac{R^2}{r^2}\sin^2\theta.
\label{e391}
\end{equation}
This implies that the energy flux is concentrated near the equatorial
plane. This $\theta$-dependence of the energy flux is used by many authors~\citep{bgh1,
kl03}. On the other hand, at large distances $r \gg R_{\rm L}$, \citet{ingr} and \citet{mich74} found another
asymptotically radial solution, with $E_{\theta}(\theta) = B_{\varphi}(\theta)$, 
resulting in a radial Poynting vector with arbitrary $\theta$-dependence.  
In all of these solutions, the relation
\begin{equation}
S(\theta) \propto B_{r}^2(\theta)\sin^2\theta
\label{e392}
\end{equation}
is valid.

\citet{bg99} generalized the split-monopole model, showing that in the force-free approximation the ``inclined split
monopole field'' is a solution of the problem as well.
In this solution,
\begin{eqnarray}
B_{\varphi} & = & E_{\theta}
= - B_0\left(\frac{\Omega R}{c}\right)\frac{R}{r}\sin\theta \,{\rm sign}{\rm \Theta}
\label{e38''}
\end{eqnarray}
and $B_{\rm p} = B_0(R/r)^2 \,{\rm sign}{\rm \Theta}$, where
\begin{equation}
{\rm \Theta} = \sin\chi\sin\theta\sin(\varphi - \Omega t +\Omega r/c)+ \cos\chi\cos\theta.
\label{Theta}
\end{equation}
In this case, within the cones
~\mbox{$\theta < \pi/2 - \chi$}, $\pi - \theta < \pi/2 - \chi$ around the rotation
axes, the electromagnetic field is not time dependent; whereas in the equatorial
region, the electromagnetic fields change the sign at the instant ${\rm \Theta} = 0$. 
In other words, the condition ${\rm \Theta} = 0$ defines the location of the current sheet. 
We stress that the expression \eqref{Theta} for the shape of the current sheet remains 
true for the other radial asymptotic solutions, with $E_{\theta}(\theta) = B_{\varphi}(\theta)$ 
but arbitrary $\theta$-dependence, as well~\citep{Arz15a}. Numerical simulations obtained recently for 
the oblique force-free rotator confirm this conclusion as well~\citep{Spi06, contop1, contop2, Tch13, Phi13}.

\subsection{Magnetohydrodynamic Models}

As was already stressed, recently numerical simulations have become possible that can
simulate the structure of plasma-filled magnetospheres from first
principles. \citet*{ckf99} found an iterative way to do this and obtained the
first solution for an \emph{aligned} force-free pulsar magnetosphere
that extended out to infinity (see Figure~\ref{fig:04}). Their results were subsequently
verified by other groups within force-free and magnetohydrodynamic
(MHD) approximations (e.g.,
\citealt*{gruz1,tim06,mck06,k06,par12phaedra,2014PhRvD..89h4045R})
as well as using particle-in-cell (PIC) approach
\citep*{2014ApJ...785L..33P,ChenPIC,2014arXiv1410.3757C,2014arXiv1412.2819B}. %

\begin{figure}
  \centering
  \includegraphics[width=0.8\textwidth]{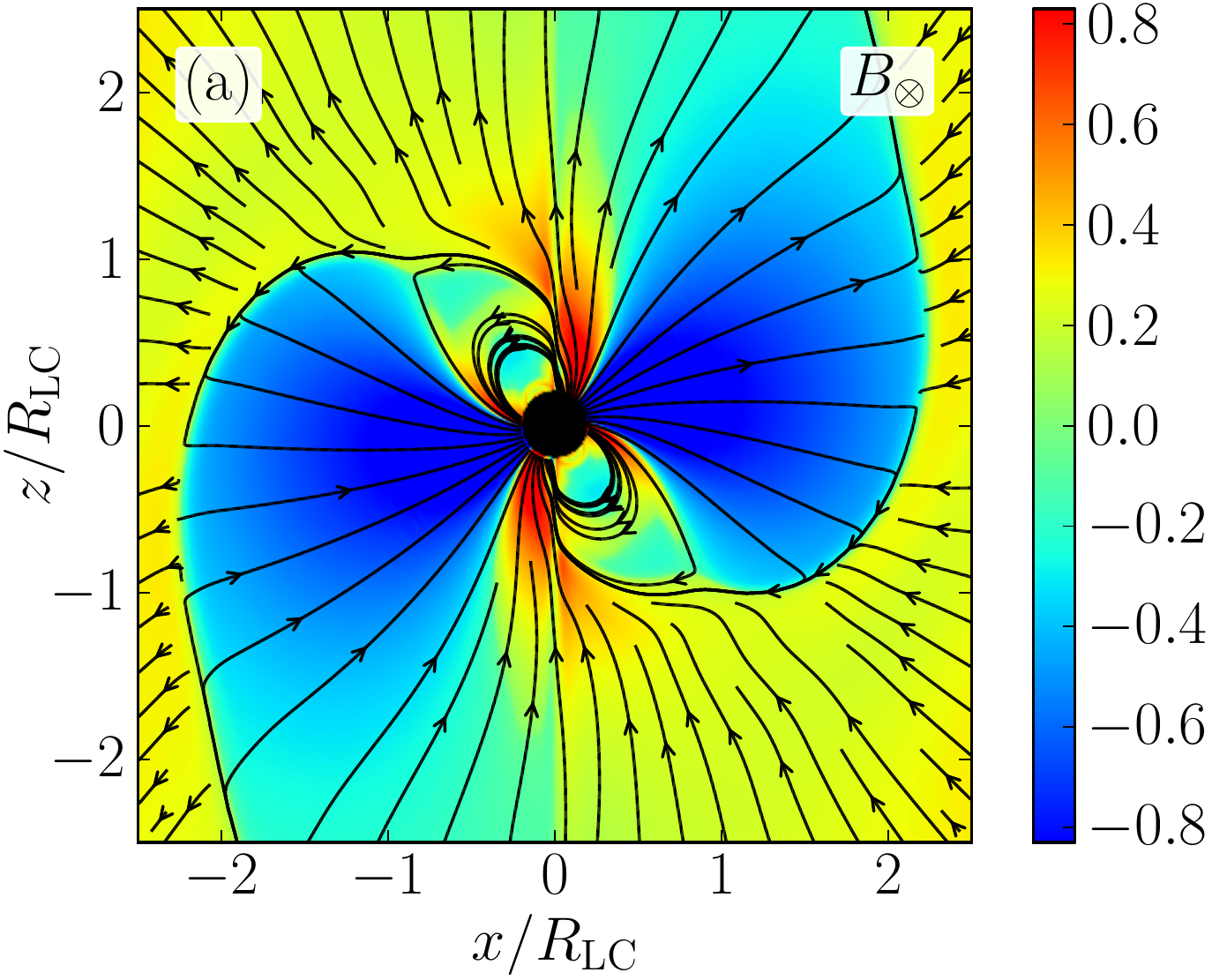}
  \caption{Slice through the $\vec m{-}\vec\Omega$ plane of a
    relativistic MHD simulation of an oblique pulsar magnetosphere
    (obliquity angle $\chi=60^\circ$) taken after $3$ rotations. Solid lines show
    field lines as traced in the image plane. Into-the--plane magnetic
    field component, $B_\otimes$, is shown with color (red -- into plane, blue
    -- out of plane). Taken from \cite{Tch13}.}
  \label{fig:3dmhd}
\end{figure}

\citet{Spi06} carried out the first 3D, oblique pulsar magnetosphere
simulations. Using the force-free approximation, he found that pulsar
spindown luminosity increases with increasing obliquity angle,
$\chi$, which is the angle between the rotational and magnetic
axes. 
The spindown obtained in such force-free and MHD models is
well-described by
\begin{equation}
  \label{eq:Lmhd}
  W_{\rm tot} = W_{\rm aligned}(1+\sin^2\chi),
\end{equation}
where $W_{\rm aligned} = m^2\Omega^4/c^3$ is the spindown luminosity
of an aligned plasma-filled pulsar magnetosphere, and $m = B_0 R^{3}/2$ is the
magnetic dipole moment of the pulsar.
More recently, these results were confirmed using time-dependent
3D force-free \citep*{contop1,petri12a,kalap12}, MHD \citep*{Tch13}, and
PIC \citep*{2014arXiv1412.0673P} studies.  Figure~\ref{fig:3dmhd}
shows a vertical slice through the results of a 3D MHD simulation of
an oblique pulsar magnetosphere with obliquity angle $\chi =
60^\circ$. One can clearly see the closed zone that extends out to the
light cylinder located at $|x| = R_{\rm LC}$, beyond which starts
a warped magnetospheric current sheet, across which all field
components undergo a jump. The structure of this current sheet is
presently poorly understood, in particular it is not known if in the
perfect conductivity limit the magnitude of the magnetic field in the
sheet vanishes.

What causes this increase of spindown luminosity with the increase of
pulsar obliquity?  It turns out that there are two factors: (i) an
increase in the amount of open magnetic flux, which accounts for about
$40$\% of the increase, and (ii) redistribution of open magnetic flux
toward the equatorial plane of the pulsar magnetosphere, which
accounts for the remaining $60$\%.

In fact, the spindown trend \eqref{eq:Lmhd} can be reproduced via a
simple toy model.  Suppose that the magnetic field that reaches the
light cylinder in an oblique rotator, with inclination angle $\chi$,
retains the dipolar structure at $r=r_0\gg R_{\rm LC}$, 
\begin{equation}
B_r = \frac{B_0  r_0^2}{r^2} \sin\theta_{\rm m}, \label{eq:Brmhd}
\end{equation}
where
$\theta_{\rm m}$ is the magnetic colatitude, or the angle away from
the magnetic axis,
\begin{equation}
\label{eq:thetamdef}
\theta_{\rm m} = \arccos(\sin\chi\sin\theta\cos\varphi+\cos\theta\cos\chi).
\end{equation}
How would the
pulsar spindown change if we kept the total open magnetic flux,
$\Phi_{\rm open} = \pi r_0^2 B_0$, fixed, and inclined the pulsar,
i.e., increased $\chi$?  
To find this out, let us first compute the angular distribution of $\varphi$-averaged $B_r^2$:
\begin{equation}
\langle B_r^2\rangle_\varphi = B_0^2(0.5\sin^2\chi
\sin^2\theta + \cos^2 \theta \cos^2 \chi).
\label{eq:Brsq}
\end{equation}
Now, making use of the fact that $B_\varphi \approx E_\theta = -B_r\Omega
r\sin\theta/c$ and the radial Poynting flux is $S_r = cE_\theta
B_\varphi/4\pi \approx ( B_r\Omega
r\sin\theta)^2/4\pi c$, we obtain \citep{Tch15}:
\begin{equation}
W_{\rm toy}(\chi) = \iint S_r\,{\rm d}\omega = \frac{\Omega^2\Phi_{\rm open}^2}{7.5\pi^2c} (1 + \sin^2 \chi),
\label{eq:Pdip}
\end{equation} 
where the integral is over, e.g., a sphere of radius $r_0$.
Clearly, if the total magnetic flux
$\Phi_{\rm open}$ is held constant, the nonuniformity in the surface distribution of magnetic flux
causes an \emph{enhancement in spindown losses} at higher inclination
angles, consistent with the numerical simulations (see
Eqn.~\ref{eq:Lmhd}). In the simulations, we find that the magnetic flux
itself is an increasing function of $\chi$, $\Phi_{\rm open}\propto(1+0.2\sin^2\chi)$,\footnote{This 
relation coincides exactly with one obtained by~\citet{bgi93} analytically.} so the two effects -- of the
non-uniformity of the open magnetic flux and the change in the amount
of open magnetic flux -- have very similar inclination-dependences.

In reality, the structure of the plasma-filled magnetosphere is of
course more complex than given by Eqn.~(\ref{eq:Brmhd}), but
the qualitative effect is the same: the inclination of the magnetic axis
relative to the rotational axis leads to the shift of the peak of
$|B_r|$ away from the axis and toward the equatorial
plane and an increase in the spindown luminosity \citep{Tch15}.

More recently, PIC models have been developed and, in those cases when
the magnetospheric polar cascade is efficiently operating and is able
to fill the magnetosphere with abundant plasma, are in agreement in
the amount of spindown and large-scale dissipation as in MHD
simulations (see
e.g. \citealt{2014ApJ...785L..33P,ChenPIC,2014arXiv1410.3757C,2014arXiv1412.2819B,2014arXiv1412.0673P}). Interestingly,
if a mechanism of pair formation operates only near the surface of the
star, aligned pulsar magnetospheres in PIC simulations do not reach a
force-free state \citep{ChenPIC}. In fact, PIC simulations, into which
simplified physics of the polar cascade was included, show the
development of the polar cascade and of a
force-free--like magnetosphere only for high inclinations,
$\alpha>40^\circ$ \citep{2014arXiv1412.0673P}.

\section{Observations: Energy Loss from Pulsars}\label{sec:magrotobs}

\subsection{Spindown}

In principle, the time rate of change of pulse period $\dot P$ is easy to measure.
Because individual pulses can be numbered,
period and period derivative are among the fundamental parameters of a timing model.
Period derivative is easily associated with the loss of rotational kinetic energy via
electromagnetic radiation and particle wind.
The Larmor formula for magnetic dipole radiation then directly associates
energy loss with the magnetic moment of the neutron star.
This provides a characteristic scale.

Pulsars with periods longer than a fraction of a second show timing noise: random variations
of pulse arrival time that change slowly with time \citep{Helf80}.
These variations are most extreme for the young Crab and Vela pulsars \citep{Boy72,lg-s,Scott03,Dod07}.
Among millisecond pulsars, B1937+214 shows timing noise, but other millisecond pulsars may not \citep{Kas94,Cog95}.

Several pulsars show clear variations in spindown rate associated with changes in pulse properties.
The radio pulsars B1931+24, J1832+0029, and J1841-0500
intermittently switch between an ``on'' radio-loud state in which they appear as ordinary radio
pulsars, and an ``off'' state in which no radio emission is detected.
The spin-down rate is higher in the ``on'' state than the ``off'' state,
by a factor of
$f_{{\rm on}\rightarrow{\rm off}} = \dot P_{\rm on}/ \dot P_{\rm off} \approx 1.5$ for B1931+24 \citep{Kra06}
and J1832+0029 \citep{Lyn09},
and $f_{{\rm on}\rightarrow{\rm off}} \approx 2.5$ for J1841$-$0500 \citep{Cam12}.
The gamma-ray
pulsar J2021+4026 \cite{All13} displays two states with intensities different by 20\% and with distinct pulse profiles,
each associated with a different spindown rate:
$f_{{\rm on}\rightarrow{\rm off}} = 1.04$.
Pulsar B0919+06 shows quasiperiodic variations between two states with different spindown rates and different pulse profiles
\citep{Per15}.
\citet{Lyn10} propose that the phenomenon of intermittency is quite general:
they find that timing noise for six pulsars can be expressed as the superposition of two states,
characterized by distinct pulse profiles and spindown rates,
with rather rapid changes between states.
From these discussions it is clear that magnetospheric structure affects spindown,
as one would suspect from theoretical considerations discussed in Section 2.3 above.

\citet{Kra06} \citep[see also][]{bn2007} proposed that two distinct magnetospheric states
lead to the observed difference in spindown rates.
They associated the ``off'' state with a magnetosphere depleted of charge,
and the ``on'' state with magnetospheric currents sufficient to produce the observed change in spindown.
\citet{Li12} observe that the simplest model for the ``on'' state is the force-free magnetosphere \citep{Spi06},
which exhibits spindown rates at least three times that of a vacuum dipole.
They suggest a modified picture where the ``on'' state is the force-free magnetosphere,
and the ``off'' state has no charge on open field lines, but carries
the Goldreich-Julian charge on closed field lines.
This leads to ratios $f_{{\rm on}\rightarrow{\rm off}}=1.2$ to $2.9$ for inclination angles of $\chi>30^{\circ}$.
Smaller inclinations lead to larger $f_{{\rm on}\rightarrow{\rm off}}$.

\subsection{$\sigma$-Problem}\label{sec:sigma_prob}

Thus, we see that all analytical and numerical force-free models of the pulsar magnetosphere
demonstrate the existence of an almost-radial highly-magnetized wind, flowing outward from the pulsar
magnetosphere. On the other hand, observations show that most energy far from the neutron star
must be carried by relativistic particles~\citep{kc1, kc2}. For example, the analysis of the
emission from the Crab Nebula in the shock region located at a distance of $\sim 10^{17}$ cm
from the pulsar in the region of interaction of the pulsar wind with the supernova remnant
definitely shows that the total flux $W_{\rm em}$ of the electromagnetic energy in this region
is no more than $\sim 10^{-3}$ of the particle energy flux $W_{\rm part}$. 
Thus, in the asymptotically far region of pulsar models, the Poynting
flux must be completely converted into an
outgoing particle flux before reaching the reverse shock at distances
of $\sim0.1$~pc. 
Axisymmetric numerical models of jets from radio
pulsars are constructed exactly under this assumption \citep[][and references therein]{Kirk09}.

The transformation from Poynting flux to particles apparently occurs much closer to the neutron star,
at distances comparable to the size of the light cylinder. This is evidenced by
the detection of variable optical emission from companions in some close binary systems
involving radio pulsars
\citep{Fru88,Kulk88,Fru90,RyTa91,Stapp96,Rob11,Pall12,Rom12,Kap13,Bre13}. %
This variable optical emission with a period equal exactly
to the orbital period of the binary can be naturally related to the heating of the companion's
surface facing the radio pulsar. It was found that the energy reradiated by the companion star
almost matches the total energy emitted by the radio pulsar into the corresponding solid angle. %
Clearly, this fact cannot be understood either in the magnetic-dipole radiation model or by
assuming a Poynting-dominated strongly-magnetized outflow, since the transformation coefficient
of a low-frequency electromagnetic wave cannot be close to unity. Only if a significant fraction
of the energy is related to the relativistic particle flux can the heating of the star's surface
be effective enough. 
Moreover, eclipses of the double-pulsar system 
show effects of the particle wind from one object impinging upon \citep{Lyn04,Jen04,McLa04,Lyu04,Demo04}.
Therefore, the so-called $\sigma$-problem --- the question as to how the energy
can be converted from electromagnetic fields to particles in the pulsar wind --- remains one of
great unsolved problems of modern astrophysics. We note that the
$\sigma$-problem appears to be rather general and in addition to
neutron-star powered outflows it applies to black-hole powered, collimated
outflows known as astrophysical jets, such as in the magnetically-arrested
disk (MAD) scenario \citep{2011MNRAS.418L..79T,2012MNRAS.423L..55T,2014Natur.510..126Z,2014arXiv1410.7310Z,2014Natur.515..376G,2015ASSL..414...45T}. Theoretical models suggest that
the jets accelerate roughly up to the
equipartition between the magnetic and kinetic energies, beyond which
the acceleration slows down dramatically, locking in a substantial
fraction of energy in the magnetic form
(\citealt{Tch09,2009MNRAS.394.1182K,2010MNRAS.402..353L}; however, see
\citealt{2010NewA...15..749T}).

\section{Theory: Polarization and Refraction of Radio Emission}

\subsection{Polarization}

Pulsar emission is usually highly linearly polarized, with a small fraction of circular
polarization. Like the mean profile, the profiles in polarization states are stable
and are characteristic of the pulsar. This long-term stability of the mean properties
indicates that the pulse arises as a cut through a radiation cone, with properties that
are set by stable properties of the underlying neutron star. It is widely assumed that
the polarization is determined by magnetic fields in or above the emission region.
Those magnetic fields, in turn, are anchored in the solid crust of the neutron-star
\citep{Man95}. 

Rapid swings of the position angle of linear polarization through the pulse, first observed
in the Vela pulsar by \citet{Radetal69}, suggest that a vector 
fixed in the frame of the rotating star influences the direction of linear polarization, a geometric inference known as
the rotating-vector model. \citet{RadCoo69} proposed that this vector is the magnetic pole of
the pulsar's nearly-dipolar magnetic field; this physical interpretation is known as the
magnetic-pole model. 

Radiotelescopes can measure polarization properties of individual pulses for a number of strong pulsars.
Such studies indicate the presence of orthogonal
modes, with polarization differing by $90^\circ$, and intensities varying from pulse to pulse \citep{mth75,brc76,crb78,br80,Sti84a,Sti84b,mck03a}. 
For most pulsars, present radiotelescopes can determine only average polarization properties; 
nevertheless the presence of two competing orthogonal modes can explain the observed departures from the characteristic pattern, for most of these weaker pulsars. 
Thus, the rotating-vector model, with two orthogonal linearly-polarized modes, successfully describes the characteristic
swing of the angle of polarization with pulse phase for most pulsars, across a wide range of pulsar parameters and despite observational selection effects~\citep{rk83,Ran86,Lyn88}.
The
two modes are usually interpreted as the X-mode, with wave electric field perpendicular to stellar
magnetic field ($E_W \perp B_0$); and the O-mode, with a component of electric field parallel to
stellar field ($E_W || B_0$). %
Both modes appear to be present, at some level, for all radio pulsars.

Work to model the polarization properties of pulsars in more detail, including the
circular-polarized profile, have led to mapping of the polarization properties on
the Poincar\'e sphere describing the Stokes parameters~\citep{McK09,Chu11,Chu11b}.
These show a rich variety of patterns, with greater modulation of polarization being
indicative of more complex patterns. Analysis of these patterns suggest emission, or
refractive scattering within the pulsar's light cylinder. For some pulsars, the emission,
or reprocessing region is inferred to lie at altitudes of 10 to 40\% of the light-cylinder
radius.

\subsection{Rotating Vector Model}

\begin{figure}
  \centering
  \includegraphics[width=0.6\textwidth]{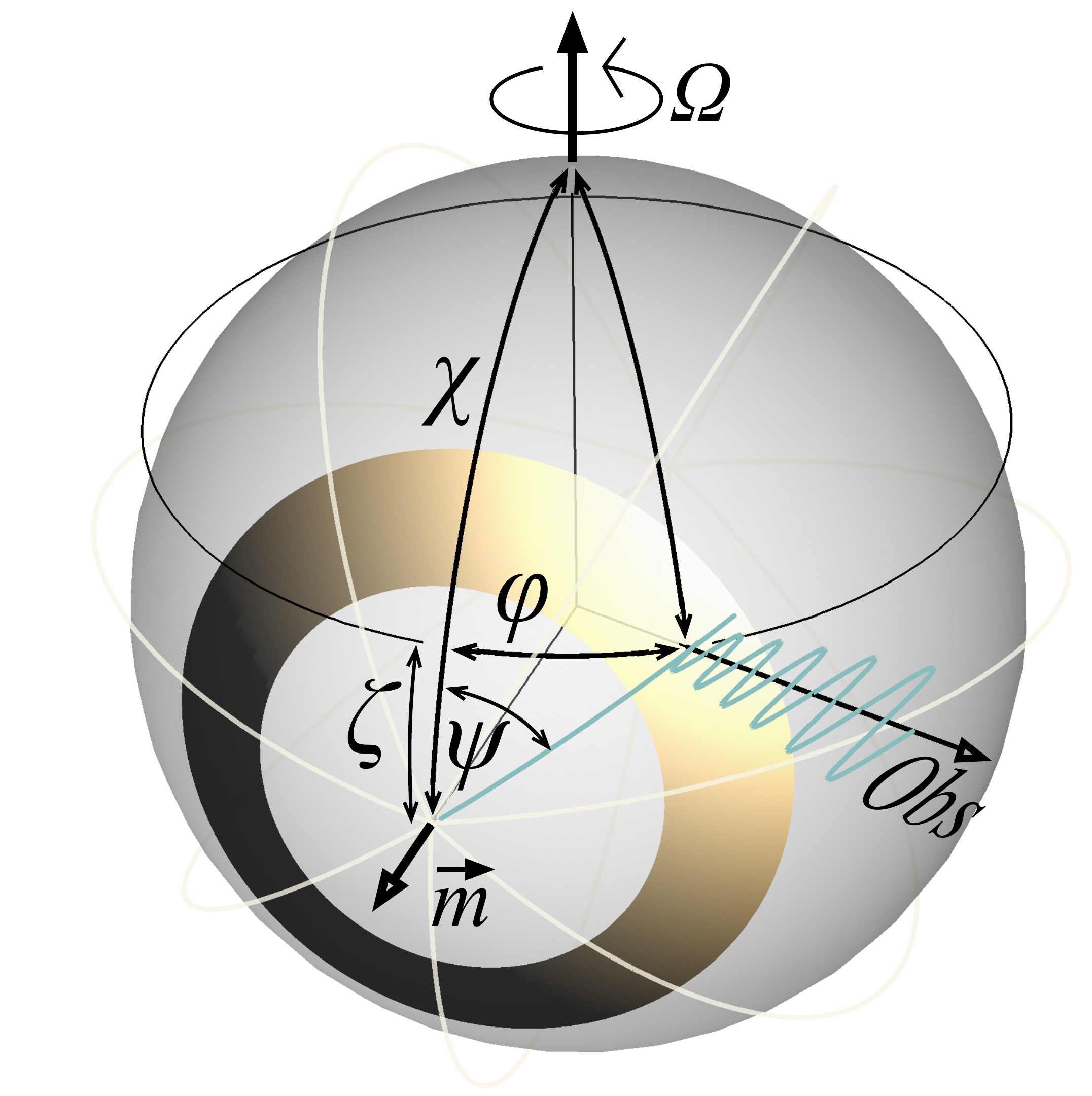}
  \caption{Geometry for the rotating vector model. The emission surface is shown as a golden band.
  As the pulsar rotates, the line of sight to the observer traverses a circle of constant latitude, producing an observed pulse each time the emission surface crosses the line of sight.
  The dipolar magnetic field, projected onto the radius of the emission surface, radiates from the magnetic pole $\vec m$.
  The instantaneous polarization of the observed radiation is parallel to the field lines, at position angle $\psi$ at the observer.
  \label{fig:RotatingVector}}
\end{figure}

The standard relation for the rotating vector model
describes variation of the position angle of polarization $\psi$ in the mean profile, under the assumption that the
the hollow-cone model is valid.  
In other words, it assumes that 
all absorption is absent, and that the 
magnetic field is dipolar in the emission region, precisely where the polarization is determined.
This relation takes the form:
\begin{equation}
\psi = {\rm arctan} \left(\frac{\sin \chi \sin \phi}
{\sin \chi \cos \zeta \cos \phi - \sin \zeta \cos \chi}\right),
\label{p.a.}
\end{equation}
Here, once again, $\chi$ is the inclination angle of the magnetic dipole to the rotation axis,
$\zeta$ is the angle between the rotation axis and the direction toward the observer, and $\phi$
is the phase of the pulse. Figure\ \ref{fig:RotatingVector} illustrates the geometry. 
Equation  (\ref{p.a.}) has been used for many years 
in estimating the pulsar inclination angle, which  is a very important parameter for 
determining the structure of the magnetosphere. Aberration and retardation effects~\citep{BCW}
have been included in only some studies~\citep{ ML2004, Krzes}.

The rotating vector model, extended further with the ``hollow cone''
model, is based on the following three assumptions (see, e.g.,~\citet{ManTay}): the
formation of polarization occurs at the point of emission; radio waves propagate along
straight lines; and cyclotron absorption can be neglected. But all these assumptions turn
out to be incorrect. \citet{Bar86} showed that in the innermost regions of the
magnetosphere, the refraction of one of the normal modes is significant. After publication
of the work of Mikhailovskii's group \citep{Mikh82}, it became clear that cyclotron
absorption can significantly affect the radio emission intensity. The influence of the
magnetosphere plasma on variation of the polarization of radio emission propagating in
the internal regions of the magnetosphere also must not be neglected~\citep{PetrovaLyu}.

The ``limiting polarization'' is the most important effect of magnetospheric propagation.
Radio emission in the region of dense plasma consists of a superposition of normal modes: 
in particular, the principal axes of the polarization ellipse must remain aligned with the magnetic-field direction in the picture plane.
Polarization in the vacuum region is independent of magnetic field.
Hence, between the two lies a transition layer,
past which the polarization is no longer affected by the magnetospheric
plasma. For typical parameters of the pulsar magnetosphere, 
the formation of polarization occurs not at the emission point but at a distance of about
$0.1 R_{\rm L}$ from it~\citep{ChengRud, Barnard}. Taking this effect into account should
also explain the observed fraction of circular polarization of the order of (5-10)\%.
Therefore, a consistent theory of radio wave propagation
in the magnetosphere is required for a quantitative comparison of theoretical results on radio emission with
observational data.

\subsection{Propagation effects}\label{sec:propeffects}

\begin{figure}[t]
\centering
\includegraphics[scale=0.37,clip=true,trim=0.8cm 0.0cm 0.0cm 0.0cm]{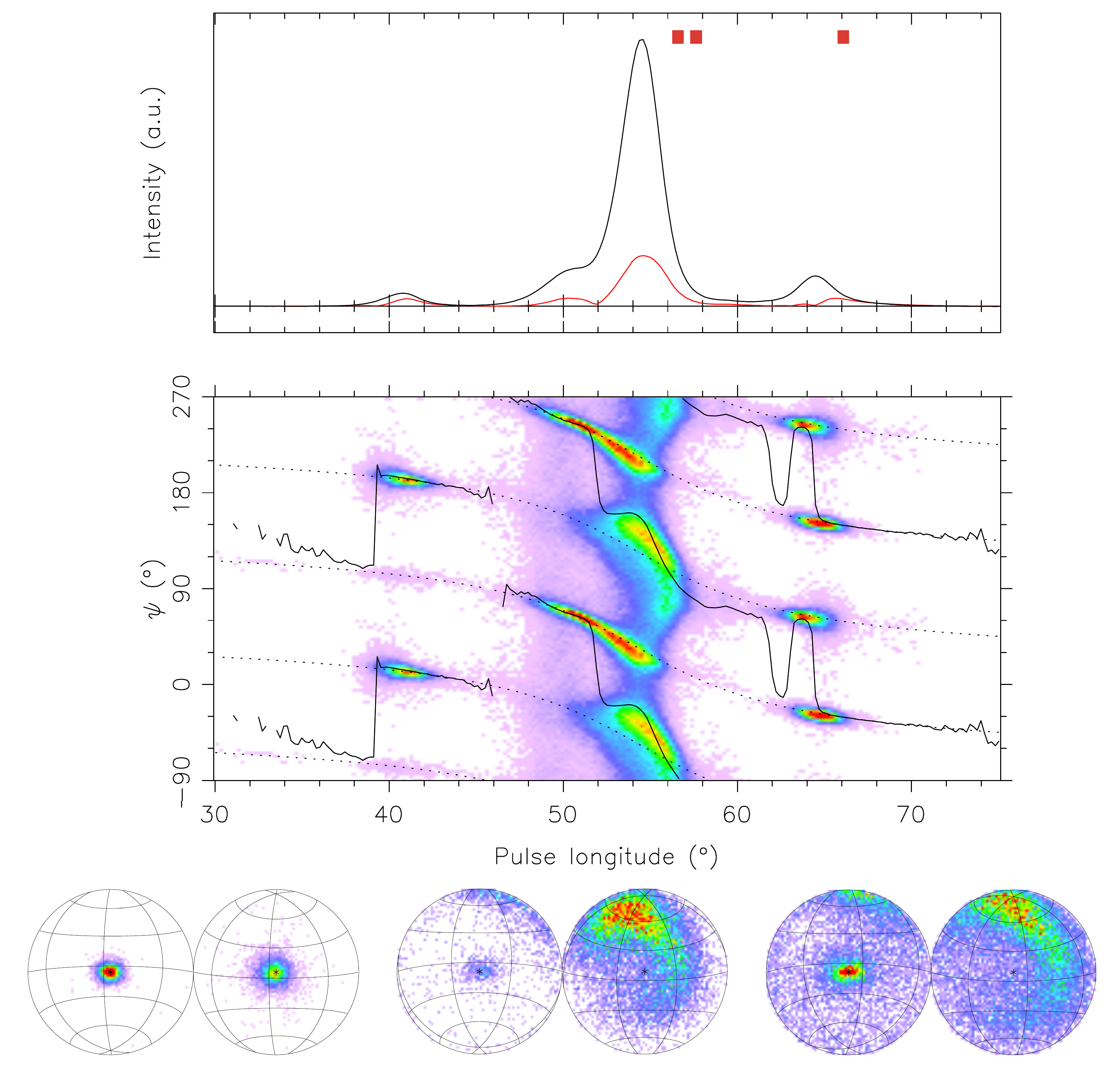}
{\caption{Polarization properties of the strong pulsar B0329$+$54. 
Upper panel: Average pulse profile, showing both total (black) and polarized (red) intensity plotted with pulse phase.
Middle panel: Histogram of angle of linear polarization $\psi$ plotted with pulse phase. Dotted curves show polarization for the rotating-vector model.  Red shows most common values,
ranging through less-common green, blue, and violet, to white for not observed.
One linear polarization (presumably X) tracks the rotating-vector model well, while the other (presumably O) shows large variations in polarization
and timing.
Lower panels: Polarization parameters on the Poincar\'e sphere, for the 3 intervals of pulse phase indicated by red boxes at in the upper left panel.
Each pair of disks shows 2 hemispheres.  Linear polarization lies on the equator; pure right circular polarization is at the
upper pole, pure left circular at the lower.
This image from \citet{edwsta04} is reproduced by courtesy of B. W. Stappers.
        }\label{fig:0329pol}}
        \end{figure}

At present, the theory of radio wave propagation in the magnetosphere of a pulsar can be
considered to provide the necessary precision~\citep{Petrova, Wang, BesPhil, BesPhil2012,
KravtsovOrlov}. Four normal modes exist in the magnetosphere~\citep{bgi93, lg-s}. Two of
them are plasma modes and two are electromagnetic, which are capable of departing from
the magnetosphere. An extraordinary wave (the X-mode) with the polarization perpendicular
to the magnetic field in the picture plane propagates along a straight line, while an
ordinary wave (the O-mode) undergoes refraction and deviates from the magnetic axis.
An important point here is that for typical magnetosphere parameters, refraction occurs at
distances up to $0.1 R_{\rm L}$, i.e., it can be considered separately from the cyclotron
absorption and the limiting polarization. As shown in Figure\ \ref{fig:0329pol}, the pulsar B0329$+$54 shows both X- and O-modes,
with the O-mode displaying deviations from the rotating-vector model because of refraction\ \citep{edwsta04}.

Based on the~\citet{KravtsovOrlov} method, \citet{BesPhil2012} have used such a
theory of wave propagation in a realistic pulsar magnetosphere, taking corrections to the
dipole magnetosphere into account (based on the results obtained by numerical simulation
in~\citet{Spi06}), together with the drift of plasma particles in crossed electric and magnetic
fields, and a realistic particle distribution function. The theory developed allows dealing
with an arbitrary profile of the spatial plasma distribution, which may differ from the one in
the hollow-cone model, because precisely the inhomogeneous plasma distribution leads to the
characteristic `patchy' directivity pattern~\citep{rk90}.

The main result consists in the prediction of a correlation between the sign of the circular
polarization (the Stokes parameter $V$) and the sign of the derivative of the change in
the polarization of the position angle, $\psi$, along the profile, ${\rm d} \psi/{\rm d}\phi$,
where $\phi$ is the phase of the radio pulse. For the ordinary mode, these signs must be opposite
to each other, while for the extraordinary mode, they must coincide. 
Figure\ \ref{fig:0329pol} shows this pattern as well.
As was noted, refraction
of the ordinary wave leads to a deviation of beams from the rotation axis, and therefore the
ordinary wave pattern should be broader than for the extraordinary wave. In the case of the
ordinary mode, double radio emission profiles should mainly be observed, while single
profiles should be observed in the case of the narrower extraordinary mode~\citep{bgi93}.

As was shown by~\citet{And10}, observations fully confirm the prediction of correlation between signs of $V$ and ${\rm d} \psi/{\rm d}\phi$. 
The analysis used over 70 pulsars
with well-traced variation of the position angle and the sign of the
circular polarization $V$,  
chosen from reviews of pulse profiles~\citet{WeltevJohns, HankinRankin}.
Table\ \ref{tab:1} presents the results of the analysis.
Pulsars with opposite signs of the derivative
${\rm d} \psi/{\rm d}\phi$ and the Stokes parameter $V$ were placed
in class O, while those with identical signs were placed in class X. As can
be seen from the Table, most of the pulsars exhibiting a double-peaked (index D)
profile indeed correspond to the ordinary wave, while most of the pulsars with
single-peaked profiles (index S) correspond to the extraordinary wave. Moreover,
the average width of the radiation pattern 
for ${\rm O_{D}}$ pulsars is indeed
about two times larger than the average width of the radiation pattern for ${\rm X_{S}}$
pulsars. 
For the pulse width, the analysis used 
the width at the 50\% intensity level $W_{50}$, 
normalized to the pulsar period $P$.
The existence of a certain number of pulsars of classes ${\rm O_{D}}$ and
${\rm X_{S}}$ should not give rise to surprise, because for central passage through
the directivity pattern, independently of whether it corresponds to the O-mode or
to the X-mode, a double-peaked profile should be observed, while for lateral passage,
a single-peaked profile should be observed.

\begin{table}
\caption{Statistics of pulsars with known circular polarization V and
variation of position angle $\psi$.}
\label{tab:1}       %
\begin{tabular}{ l c c c c c}
\hline\noalign{\smallskip}
{\bf Polarization Mode} & \multicolumn{2}{c}{O}& & \multicolumn{2}{c}{X}  \\
\cline{2-3}\cline{5-6}
{\bf Profile Type}  & Single & Double && Single & Double \\
{\bf Class}  & O$_{\rm S}$ & O$_{\rm D}$ && X$_{\rm S}$ & X$_{\rm D}$ \\
{\bf Number of Pulsars}  & 6 & 23 && 45 & 6\\
{\bf Normalized Pulse Width}$^a$  &6.8$\pm$ 3.1&10.7$\pm$ 4.5&&6.5$\pm$ 2.9&5.3$\pm$ 3.0\\
\noalign{\smallskip}\hline
\multicolumn{6}{l}{ $a$\ Normalized pulse width given as: $\sqrt{P}W_{50}$\ (s$^{1/2}$ deg) }
\end{tabular}
\end{table}

Accurately taking propagation effects into account, 
\citet{And10, BesPhil} showed that such a variation of the position angle can be realized only
under conditions of low plasma density or high mean particle energy. They found significant deviations from
the standard relation of the rotating vector model (Eq. \ref{p.a.}) were obtained in the case of quite reasonable parameters
that satisfy models for particle production: for example,
a multiplicity $n_{\rm e}/n_{\rm GJ} \sim 10^4$ and an average Lorentz factor 
$\gamma \sim 50$.

\section{Observations: Polarization and Pulsar Size}

\subsection{Pulsar Emission Region Size and Shift}\label{sec:size_measurement}

The radio emission regions of pulsars lie within the light cylinder, and so have angular sizes from Earth of nanoarcseconds or less.
Resolving such an angle at radio wavelengths requires an instrument with an aperture approaching an AU,
beyond the capabilities of even the longest VLBI baselines \citep{Kard13}.
However, radio-wave scattering by the dilute, turbulent interstellar plasma yields 
the required angular resolution and offer some of the information provided by a lens of that aperture.

Interstellar scattering affects almost all astrophysical sources
at decimeter wavelengths, and for
many at shorter wavelengths.
For most radio pulsar observations, scattering is ``strong'' in the sense that paths contributing to the electric field
measured at the observer differ in length by many wavelengths.
Hence,
these paths behave like a corrupt lens \citep{Gwi98}.
The angular extent on the sky of these paths, $\theta$, delineates the ``scattering disk''.
The scale of variation of the impulse-response function at the observer, $S_{\rm ISS}$, is 
the diffractive spot size of that aperture, $\sim \lambda/\theta$.
(Here, the subscript ``${\rm ISS}$'' indicates ``interstellar scattering''.)
The scattering produces a random diffraction pattern in the observer plane with lateral scale $S_{\rm ISS}$.
The effective resolution limit of the corrupt lens, at the source, is $M S_{\rm ISS}$, where the magnification factor $M=D/R$ is
equal to the distance $D$ of the scattering material from the observer, divided by its distance $R$ from the source.
Interstellar scattering does not remove information from the pulsar signal;
rather, it adds a great deal of information about the paths taken.
The challenge facing the observer is to extract the spatial information about the source from the scattered pulsar signal.

\begin{figure}
\begin{center}
  \includegraphics[scale=0.12]{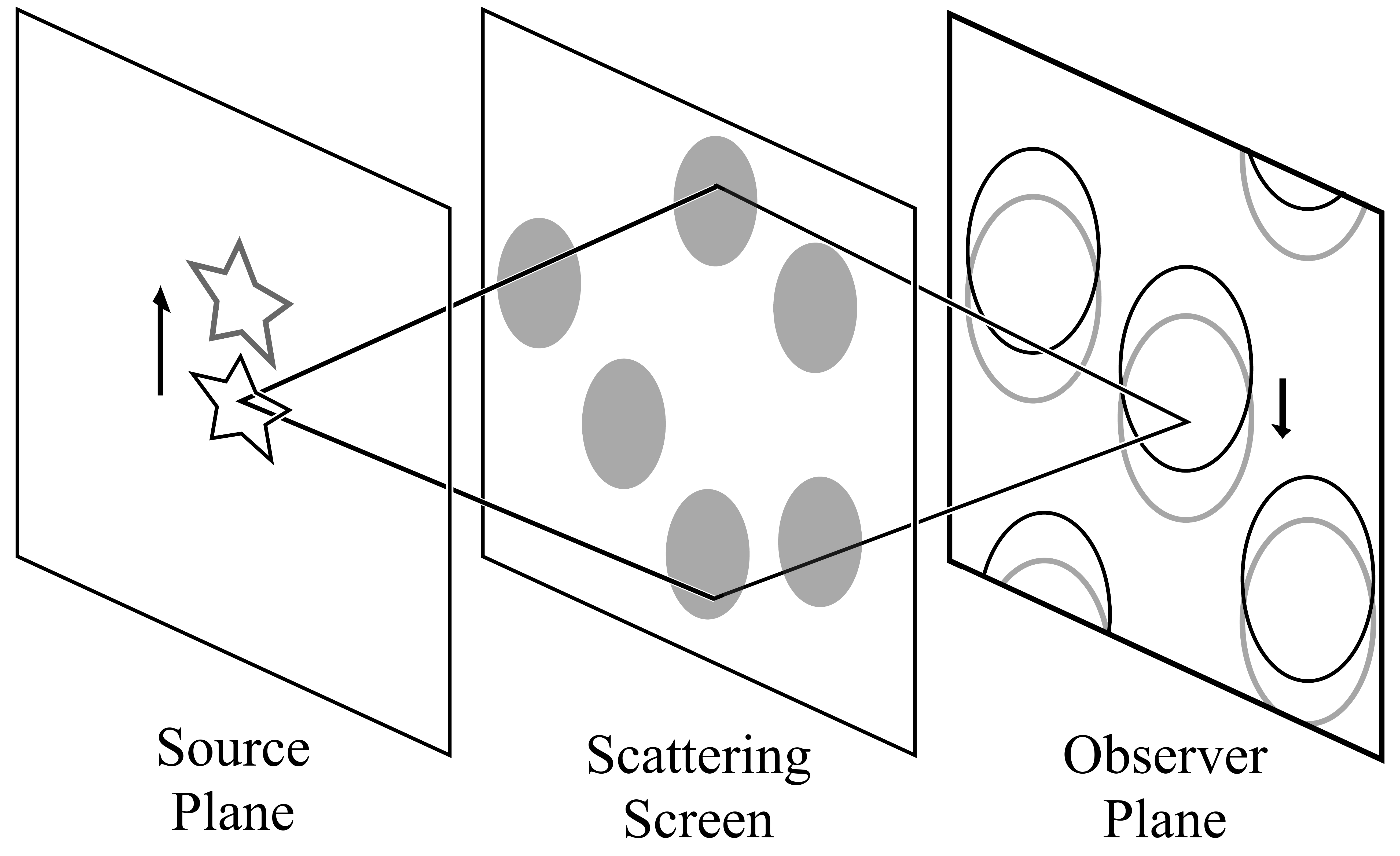}
  \end{center}
\caption{Geometry for studies of pulsar structure by scattering.  Radiation emitted by the pulsar at left travels to the scattering screen,
where fluctuations in plasma density change its phase.  The radiation arrives at the plane of the observer at right from along many paths, with different phases and
amplitudes.  Interference among paths produces a random diffraction pattern in the plane of the observer. When the source shifts to a different position,
because of either proper motion or a rotational shift in the location of the pulsar emission region,
the phases of the paths are modified, to produce a reflex motion of the diffraction pattern in the plane of the observer.}
\label{fig:3screens}       %
\end{figure}

Studies of the sizes of pulsar emission regions using interstellar scattering fall into two categories.
One category relies upon the fact that
if the emission point of the pulsar shifts across the pulse, the random image in the plane of the observer will undergo a reflex shift,
as illustrated in Figure\ \ref{fig:3screens}.
Proper motion causes a similar shift, but over time spans of many pulses.
Correlation of the scintillation spectrum across pulse phases
with later or earlier times yields the shift of the emission point \citep{Bac75,cor83,wol87,smi96,gup99,Pen14}.

A second category invokes the decreased modulation for scintillation of an extended source.
(``Stars twinkle, planets do not.'')
The depth of modulation reveals the size of the emission region \citep{Coh66,Rea72,Hew74,gwi12b,joh12b}.
More precisely, source size affects the distribution of flux density for a scintillating source.
In strong scattering many different paths, with lengths differing by many radians of phase, contribute to the electric field measured at a point in the observer plane.
The observer implicitly sums over these paths, so that the observed phase and amplitude have the character of a random walk.
The optics of this effect are similar to those of the reflex shift:
different parts of the source produce shifted, incoherent diffraction patterns at the observer,
who sums over them.
Thus,
finite source size affects the
distribution of intensity at one antenna, or that of correlated flux density between the two 
antennas of an interferometer, principally by shifting the lowest and highest intensities toward 
the central part of the distribution \citep{sch68,gwi01,joh12a,Joh13}. For realistic observations, 
the contributions of background noise, and of the noiselike statistics of the source itself, 
must be taken into account \citep{Gwi11,gwi12a,joh12a,Joh13}.

\subsection{Observations}

\subsubsection{Size of the Vela Pulsar's Radio Emission Region: $\lambda=18$ cm}\label{sec:18_cm_vela_size}

\begin{figure}[t!]
\vskip-5pt
\begin{minipage}[t]{0.40\linewidth}
\centering
    \includegraphics[width=\textwidth,clip=true,trim=5.8cm 7.2cm 5.8cm 7.4cm]{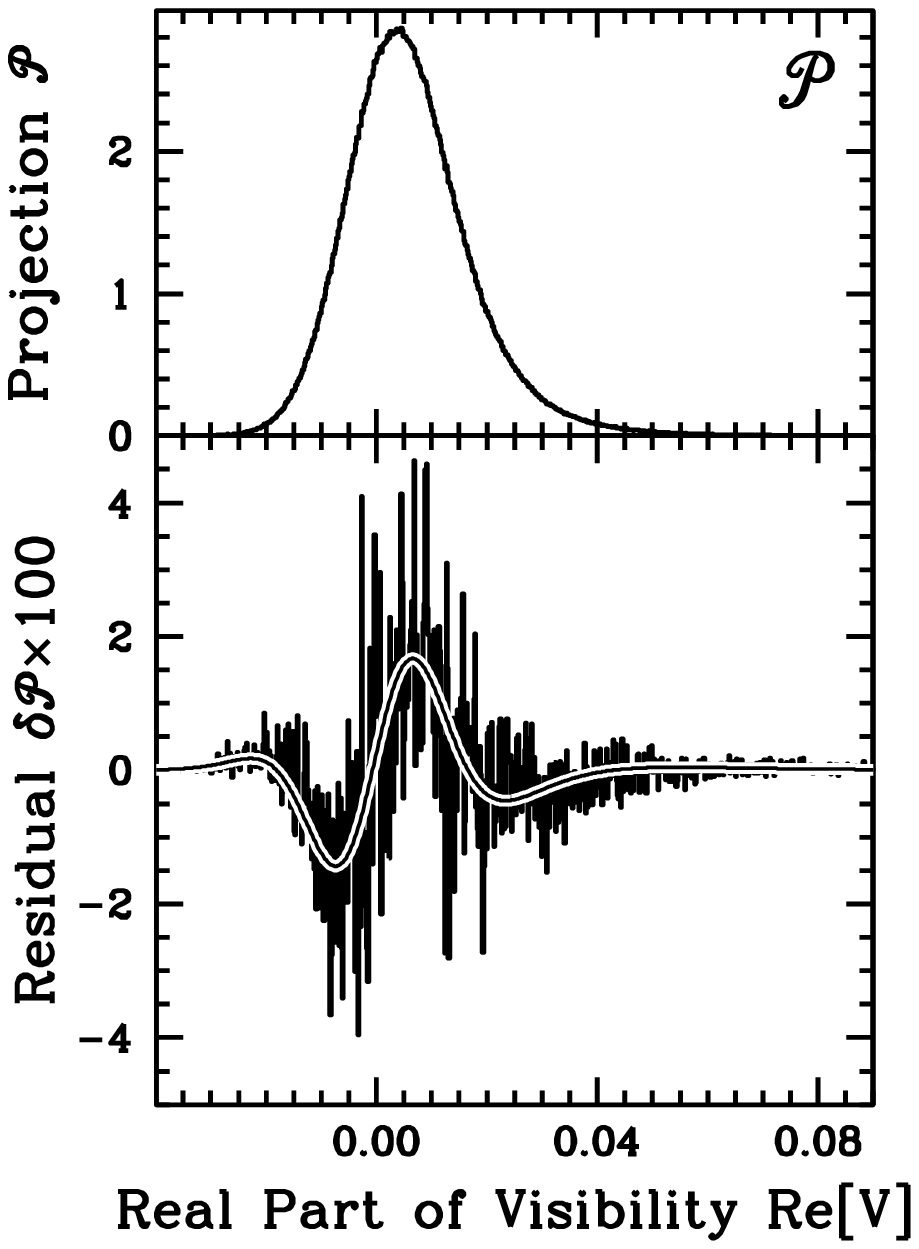}
\caption{Observed distribution of visibility projected onto the real axis ${\mathcal P}$,
in bins along the real axis.
Upper: Observed distribution. Lower: Residuals to best-fitting model with zero size for the Vela pulsar at $\lambda=18$\ cm.
Curve shows difference of finite- and zero-size models.
After \cite{gwi12b}.
\label{fig:observed_distributions}
}
\end{minipage}
\hspace{0.05\textwidth} %
\begin{minipage}[t]{0.57\linewidth}
\centering
    \includegraphics[width=\textwidth]{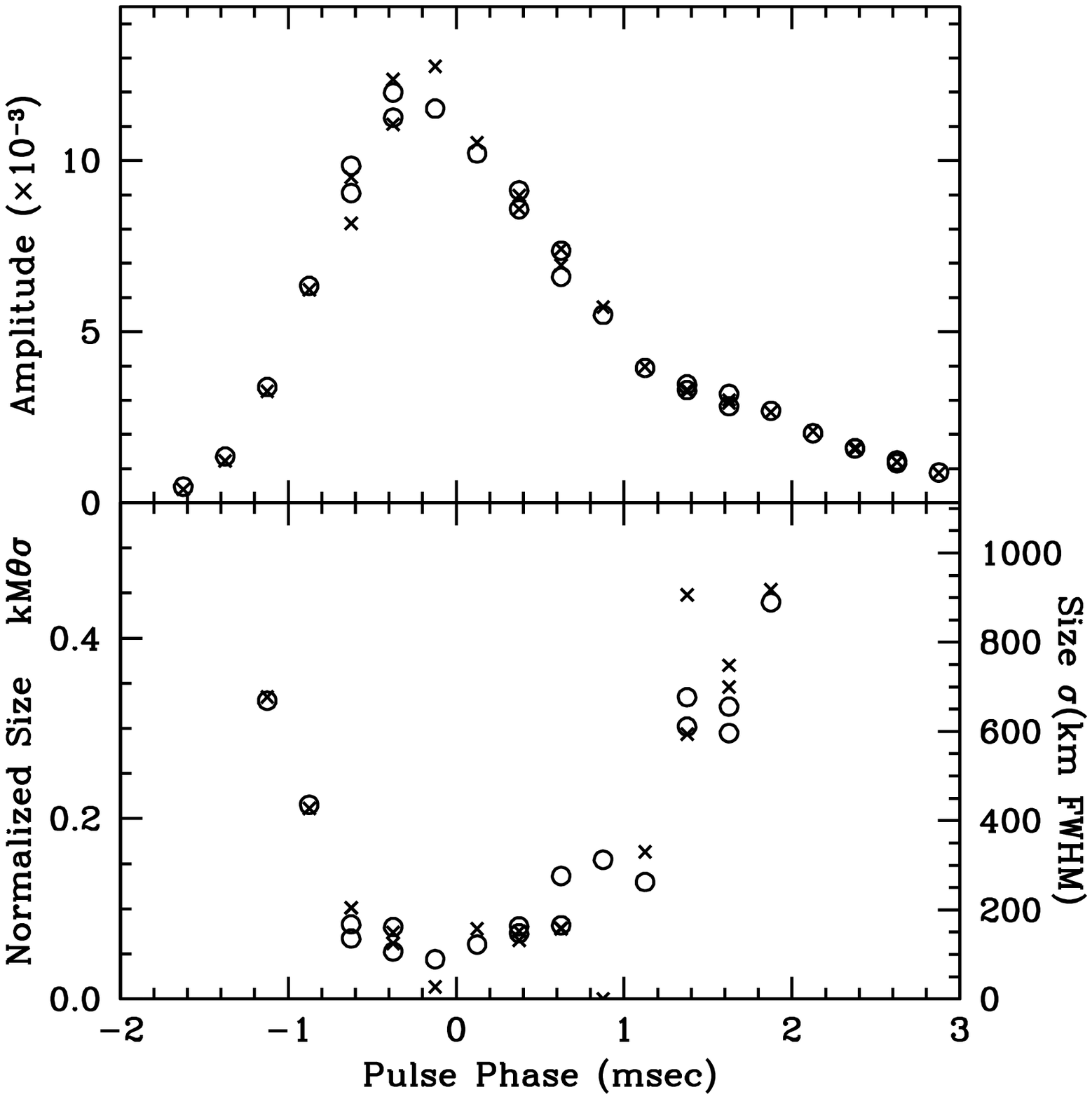}
\caption{Best-fitting amplitude (top panel) and source size $(k M \theta\sigma)$ (lower panel) plotted with pulse gate,
for 4 gates in 6 spectral ranges. 
The model for the emission region assumes a circular Gaussian distribution of emission.
After \cite{gwi12b}.
\label{fig:observed_sizes}
}
\end{minipage}
\end{figure}

The fundamental observable of interferometry is visibility, the product of electric fields at a pair of antennas \citep{tms01}.
Because electric fields of all astrophysical sources are noiselike, this product must be averaged over some range of time and frequency.
For a scintillating source, this averaging must be less than the scales of variation of the scintillation pattern with time and frequency,
to preserve the variation of visibility from scintillation \citep{gwi00}.

For a scintillating point source, in the absence of noise, the distribution of interferometric 
visibility is sharply peaked at the origin \citep{gwi01}.
The effect of a small but finite emission size is to soften the sharp peak, shift it from the origin, and narrow the distribution.
As compared with a point-source model, 
the finite-size distribution without noise peaks at larger real part,
but has lower probability density at large and small visibility,
for the same average flux density (or equivalently, the same mean visibility).

Noise broadens the distribution of visibility. Although noise blurs the distributions and their projections, the difference of point-source and finite-size distributions persists, with a characteristic W-shaped signature,
as Figure\ \ref{fig:observed_distributions} shows. 
To compare with pulsar observations, we must also incorporate the effects of intrinsic variability. Rapid variability modifies the noise statistics, while variability over longer times broadens the distribution \citep{Gwi11,gwi12a}. Consequences of these effects differ from those of emission size. 

Because finite size narrows the distribution of visibility, and noise broadens it,
the difference of best-fitting models with finite size and zero size has a characteristic W-shaped signature.
Figure\ \ref{fig:observed_distributions} shows one example of a fit for a range early in the pulse. The characteristic W-shaped residual is evident, indicating the presence of a finite emission size. A model including one additional parameter, for finite size for the pulsar's emission region, matches this residual accurately with significance exceeding 40$\sigma$. The inferred size of the emission region is 420\ km. 
From fits to gates as a function of pulse phase, we find that the 
size of the pulsar emission region is large at the beginning of the pulse, declines to near zero size near the middle
of the pulse, and then increases again to nearly 1000\ km at the end of the pulse. 
The quoted sizes indicate the full width at half maximum of an equivalent circular Gaussian model.

Theoretical models of pulsar emission typically take their starting point in the geometrical models described above.
\citet{HaBes14} made theoretical calculations of the images of pulsars as a function of pulse phase, using generic expressions for the emission altitude and beam shape.
They include effects of refraction by the magnetospheric plasma, and investigate emission heights up to 100$\times$ the radius of the neutron star.
They find a characteristic U-shaped curve of the form seen in Figure\ \ref{fig:observed_sizes}. This form results from the greater curvature of field lines further from the magnetic pole, and the consequently greater set of loci that can emit in a given direction.
Interestingly, they find that the size of the emission region is much larger than its shift over the course of a pulse.
\citet{YuMe} investigate a similar model, and find that the shift of the emission region over a pulse is indeed small.  
They suggest from geometrical arguments that emission arises at altitudes of more than 10\% of the light-cylinder radius.
\citet{Lyu99} comes to similar conclusions based on emission physics.

\subsubsection{Size of the Vela Pulsar's Emission Region at $\lambda=40$\ cm from Nyquist-Limited Statistics}\label{sec:vela_nyquist_size_measurement}

\begin{figure}
    \includegraphics[scale=0.27]{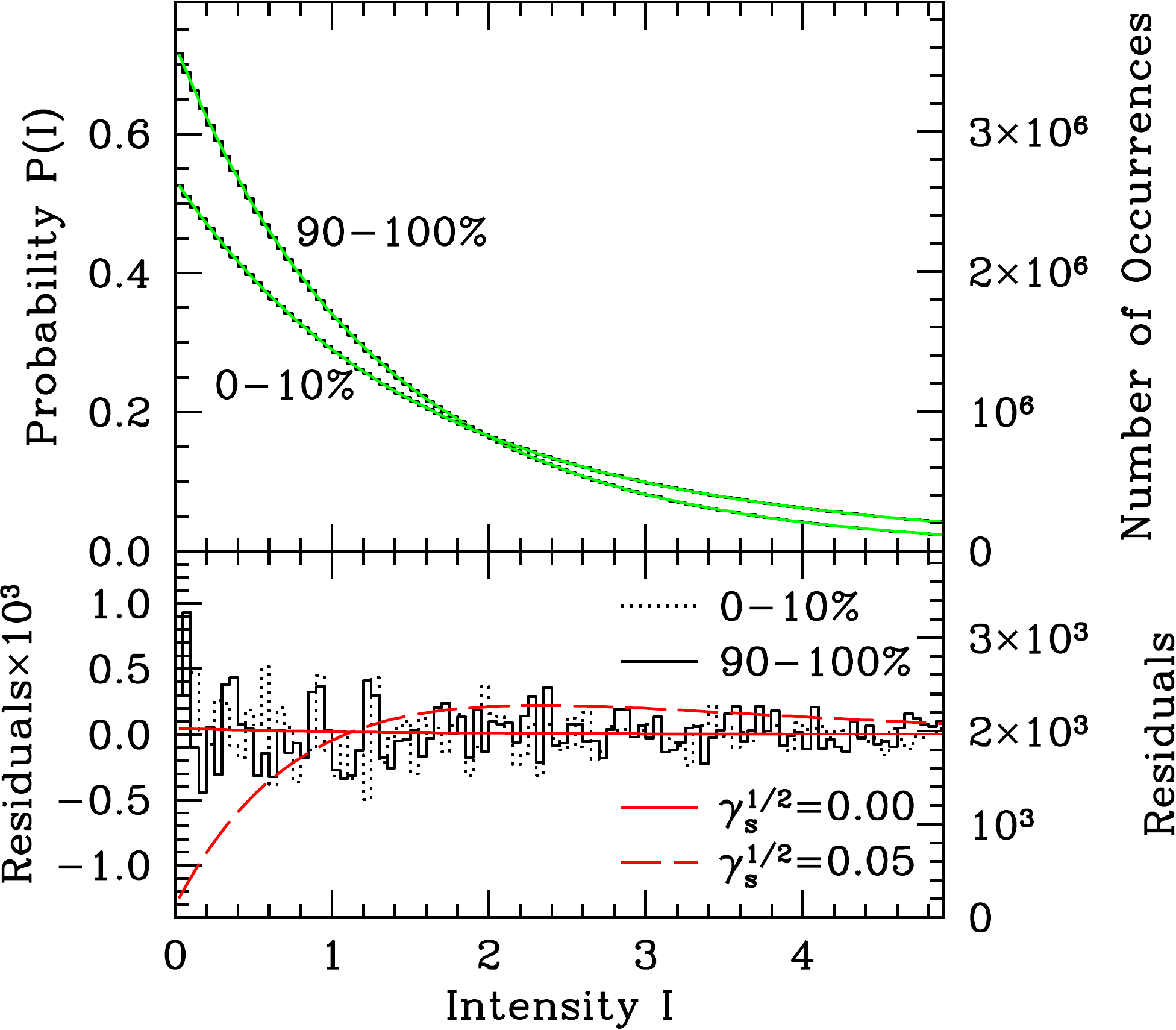}
   \includegraphics[scale=0.29]{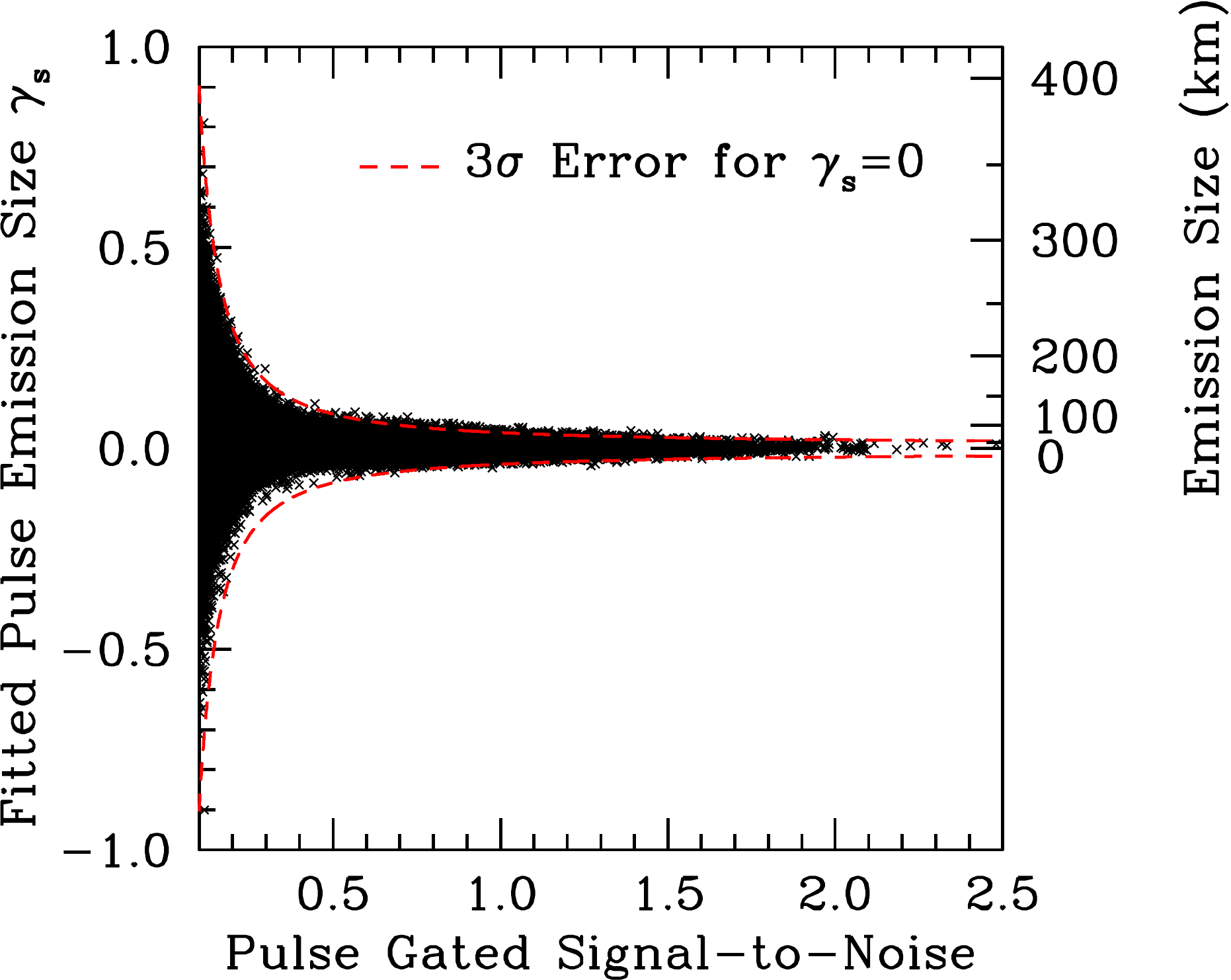}
\parbox{50mm}{\caption{Observed and model PDFs of intensity for the subsets of pulses in the top and bottom decile by pulse intensity. 
Theoretical residual curves are completely determined by a single parameter, the source size. Plotted results correspond to a point source and a source that extends over 20 km; the latter is clearly inconsistent with the observed statistics.
From \cite{joh12b}.}
\label{fig:40_cm_size}}
\hfill\parbox{60mm}
{\caption{Inferred emission sizes of individual pulses at $\lambda=40$\ cm.
The size is displayed as a function of the single-pulse
signal-to-noise ratios $S$; both linear polarizations are plotted. Because $S$
determines the standard error for each measurement to excellent accuracy, we
omit error bars and instead show the expected $\pm 3 \sigma$ errors about $\gamma_s = 0$. We do
not obtain a statistically significant detection of emission size for any pulse. From \cite{joh12b}.} \label{fig:singlepulse_size}}      %
\end{figure}

The unique nature of pulsar emission allows an elegant solution to 
determination of the distribution of intensity of a variable, scintillating source:  the formation of spectra that contain all single-pulse power.
Such spectra require a Fourier transform of a data stream that spans the entire pulse, including any scatter-broadening.
Without any averaging at all (that is, at the Nyquist limit of the data stream), such spectra show the influence of finite source size.
\citet{joh12a} calculated the distribution of Nyquist-sampled spectra, for scintillating sources with and without effects of size,
including the effects of averaging and temporal decorrelation.
With knowledge of background noise from off-pulse spectra,
the intensities of individual pulses,
and the scintillation timescale,
these statistics provide a measure of source size.
A great strength of this technique is that it can measure size 
for individual pulses,
or narrow classes of pulses.

\citet{joh12b} used the Nyquist-sampled technique to find the size of the Vela pulsar at 40-cm wavelength,
using baseband recording of the pulsar's electric field, at the Green Bank Telescope.
They found that the size was consistent with a pointlike source in all cases. 
The observational upper limit depended upon the set or subset of pulses analyzed.
Figure\ \ref{fig:40_cm_size} shows a typical example,
the distribution of intensity for the brightest 10\% of pulses, and for the weakest 10\%.
The distributions are normalized to the mean intensity in both cases,
so differences arise from the difference in signal-to-noise ratio.
The size is expressed in terms of the characteristic scales of interstellar scattering by the parameter $\gamma_s = \left( 2\pi M \theta \sigma/\lambda  \right)^2$.
Here, $\theta$ is the angular broadening by interstellar scatter, and $\sigma$ is the size of a model Gaussian distribution of intensity at the source,
both expressed as standard deviation.  The observing wavelength is $\lambda$, and $M$ is the ratio of the distance of the observer from the scattering screen, to that of the pulsar from the screen.
As the figure shows, both distributions are clearly inconsistent with a size as large as $\sigma_{\rm c} = 20$\ km,
corresponding to a full-width at half-maximum of 47\ km of an assumed circular Gaussian emission region.
Figure\ \ref{fig:singlepulse_size} shows the size of the pulsar as measured in individual pulses, with a range of signal-to-noise ratios.
From an fit to their full sample of pulses,
they obtained a 3$\sigma$ upper limit of $\sigma_{\rm c}<4$\ km (FWHM$<9$\ km).
These sizes are comparable to the size of the neutron star, and suggest a very concentrated emission region.
Theory would suggest that the shift of the emission region is still smaller \citep{HaBes14,YuMe}.

At face value, our results for the size of the emission region of the Vela pulsar at wavelengths 
of 18 and 40\ cm appear inconsistent \citep{gwi12b,joh12b}). How can the size of the emission 
region change by an order of magnitude, with a change of only 2$\times$ in observing wavelength?
Longer wavelengths are thought to arise at higher altitudes, apparently exacerbating the 
discrepancy \citep{stur71,rs}. Repeated observations have confirmed the observational results.

Refraction of a emergent double-peaked component at $\lambda=18$\ cm may be responsible.
At $\lambda=40$\ cm the pulse profile 
contains a single ``core'' component, but at  $\lambda=18$\ cm an additional, double, ``cone'' component appears \citep{kom74, ker00,joh12b}.
As discussed in Section\ \ref{sec:propeffects}, a double-peaked component indicates the presence of the O-mode,
and the effects of refraction;
whereas a single-peaked component indicates the X-mode and no refraction, and consequently a smaller size.
Magnetospheric refraction might be stronger at the shorter wavelength \cite{Aro86,Bar86,lyu00,hir01,HaBes14}.
This matches the observed pattern.

\begin{figure}[t]
\centering
\includegraphics[scale=0.43,clip=true,trim=0.0cm 7.2cm 0.0cm 4.1cm]{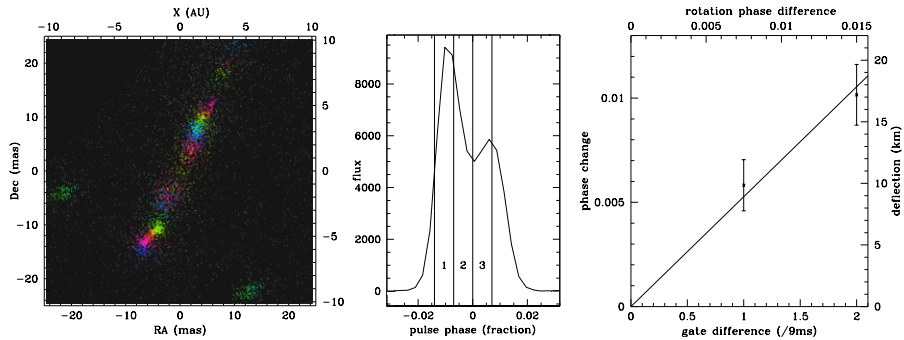}
{\caption{Left panel: Inferred image of the speckles that scatter pulsar B0834$+$06. Color is proportional to time delay, in a periodic hue map.
This image becomes
the celestial interferometer aperture, for imaging pulsar B0834$+$06.
Center: Pulse profile, with 3 bins indicated.
Right: Inferred shift of emission region with pulse phase.
This image from \citet{Pen14} is reproduced courtesy of U.-L. Pen.
        }\label{fig:femto}}
        \end{figure}

\subsubsection{Femtoarcsecond Astrometry of Pulsar B0834$+$06}

\citet{Pen14} extended the comparison of pulsar scattering patterns at different pulse phases.
They model the scattering as the interference of a set of points at the screen, the ``speckles''.
Their method isolates the wavefields of each pair of interfering speckles.
Interference of each pair of speckles acts as a 2-slit interferometer to cause the pulsar intensity observed at Earth to vary with a specific 
timescale and bandwidth. 
A shift of the position of the emission region of the pulsar, over the course of a pulse, causes a reflex shift of the interference pattern from each pair.
Pen et al. inverted the very-long baseline 
interferometry observations of this pulsar by \citet{Bri10}to infer the structure of the speckles
at the scattering screen, as shown in Figure\ \ref{fig:femto}.They use a holographic technique to partially descatter the data (see also \citet{Wal08}).  
Their technique corrects for the phase of each 
speckle relative to its neighbors, and so effectively concentrates the power and boosts the 
signal-to-noise ratio. The application of this technique to PSR\,0834+06 yields an astrometric determination 
of the phase shift across the pulse profile equivalent to an angular resolution of 150\,picoarcseconds, 
or 10\,km at the distance of the pulsar. This remarkable accuracy is comparable to the shift in
position of the pulsar due to proper motion, over a single pulse. In particular, they found that
the velocity of the radio image in the picture plane is about 1000 km s$^{-1}$, in
good agreement with theoretical prediction \citep{HaBes14}.

\section{Magnetic Axis Alignment}

\subsection{Theoretical Predictions for Motion of the Magnetic Axis}

As was shown above, resulting from the MHD theory of the neutron star magnetosphere,
the magnetospheric torque acting on the surface of a neutron star gives the positive
factor $[K_{\perp}^{A}-K_{\parallel}^{A}]$ in \eqref{8'}--\eqref{9'} corresponding to
the alignment evolution of the inclination angle. On the other hand, according to \eqref{iais},
this implies that the antisymmetric current $i_{\rm a}$ is to be large enough. E.g., for
orthogonal rotator the longitudinal current  $j$ is to be $10^3$--$10^4$ times larger
than the local Goldreich-Julian one $j_{\rm GJ}^{{\rm loc}} = {|\bf \Omega} {\bf B}|/2\pi$.
Recent simulations of pair production in the inner gap~\citep{TimArons} suggest
that the microphysics of the cascade near the polar cap can support the large currents
($j \gg j_{\rm GJ}^{{\rm loc}}$) required by the global magnetospheric structure 
(it could be accompanied by an efficient heating of the polar
cap). Similar results are obtained also by global, 3D PIC simulations
of pulsar magnetospheres \cite{2014arXiv1412.0673P}. In fact,
force-free, MHD, and PIC simulations all find that even though for an
orthogonal pulsar $j_{\rm GJ}$ essentially vanishes due to the
midplane symmetry, the magnitude of the current flowing out along the
open magnetic field lines is very similar to that of the aligned
pulsar. The results of force-free and MHD simulations tell us that \citep{Phi13},
\begin{align}
K_\parallel &= -K_{\rm aligned}\cos\chi,\\
K_\perp &= -2K_{\rm aligned}\sin\chi,
\end{align}
where $K_{\rm aligned}=W_{\rm aligned}/\Omega = \mu^2\Omega^{3}/c^3$ is
the spindown torque of an aligned rotator. Therefore,
\begin{align}
I_{\rm r}\dot{\Omega} & \approx  -K_{\rm aligned}(1+\sin^2\chi),\\
I_{\rm r}\Omega {\dot\chi} & \approx  -K_{\rm aligned}\sin\chi \cos\chi.
\end{align}
Thus, the force-free and MHD simulation results suggest that pulsars tend to become
\emph{aligned} with time. This is not surprising in the context of
previous discussion: pulsars tend to evolve toward the lowest
luminosity state, e.g., toward the aligned state (see
\eqref{eq:Lmhd}. Vacuum pulsars become aligned exponentially
fast, even before they have a chance to spin down substantially, and
generically end up with a period that is a few times their birth
period. If most pulsars were
born as millisecond rotators, this presents a problem in pulsar population synthesis
studies, as this would imply that most pulsars would have millisecond
periods, yet we observe many pulsars with periods of $\sim$second. 
In contrast, plasma-filled pulsars come into alignment much slower,
as a power-law in time, $\chi \propto t^{-1/2}$, so both the spindown and
alignment proceed at a similar rate \citep{Phi13}.

On the other hand, if there is some restriction of the value of the longitudinal current
flowing through the polar cap (no numerical simulation has such a restriction), the 
situation can be different. Such an alternative model in which both symmetric and antisymmetric 
currents correspond to the local Goldreich-Julian value was considered by \citet{bgi83,bgi93}. 
They calculated the torque associated with the Amp\'ere force arising from the interaction of the 
neutron star poloidal field with the surface currents (these currents close the longitudinal 
currents flowing in the region of open magnetosphere). One can prove by straightforward but 
cumbersome calculation that the two approaches are identical. This is a crucial assumption 
because pulsar spindown luminosity is proportional to the magnetospheric current squared.

For $i_{\rm a} \approx i_{\rm a} \approx 1$ equations \eqref{8'}--\eqref{9'} can be rewritten as
\begin{eqnarray}
I_{\rm r}\dot{\Omega}
& \approx & K_{\parallel}^{A}\cos^2\chi,
\label{8bis} \\
I_{\rm r}\Omega {\dot\chi}
& \approx & K_{\parallel}^{A}\sin\chi\cos\chi;
\label{9bis}
\end{eqnarray}
for orthogonal rotator $\cos\chi < (\Omega R/c)^{1/2}$ we have
\begin{eqnarray}
I_{\rm r}\dot{\Omega} \approx  \left(\frac{\Omega R}{c}\right) K_{\parallel}^{A}.
\end{eqnarray}
As for $\cos\chi > (\Omega R/c)^{1/2}$ evolutionary equations \eqref{8bis}--\eqref{9bis}
have an integral
\begin{equation}
\Omega \sin\chi = {\rm const},
\label{Intsin}
\end{equation}
this model predicts the evolution of the inclination angle toward an orthogonal
configuration.
Thus, two theoretical models of the neutron star
evolution give approximately identical predictions for the period derivative
$\dot P$, but opposite ones for the evolution of the inclination angle $\chi$.

\subsection{Observational Constraints on Evolution of Inclination Angle}

Measurement of the rate of change of position angle with pulse phase at the center of the pulse $d\psi/d\phi$ 
yields only a measure of the minimum angle between the line of sight and the magnetic axis, $\zeta$, 
as inspection of Eqn. (\ref{p.a.}) shows.
The angle $\zeta$ is sometimes called the ``impact angle'' (see Figure\ \ref{fig:RotatingVector}).
Consequently estimates of the inclination angle $\chi$ are indirect,
and observational tests of the theories for evolution of $\chi$ in the previous section are difficult.

As inspection of Eqn. (\ref{fig:RotatingVector}) shows, measurement of the rate of change of position angle with pulse phase at the center of the pulse $d\psi/d\phi$ 
yields only a measure of the minimum angle between the line of sight and the magnetic axis, $\zeta$, sometimes called the ``impact angle'' (see Figure\ \ref{fig:vacuumdipole}).
Consequently estimates of the inclination angle $\chi$ are indirect,
and observational tests of the theories for evolution of $\chi$ in the previous section are difficult.

\begin{figure}
\begin{minipage}[t]{0.47\linewidth} %
\centering
\includegraphics[scale=0.4,clip=true,trim=3.0cm 0.7cm 0.cm 0.cm]{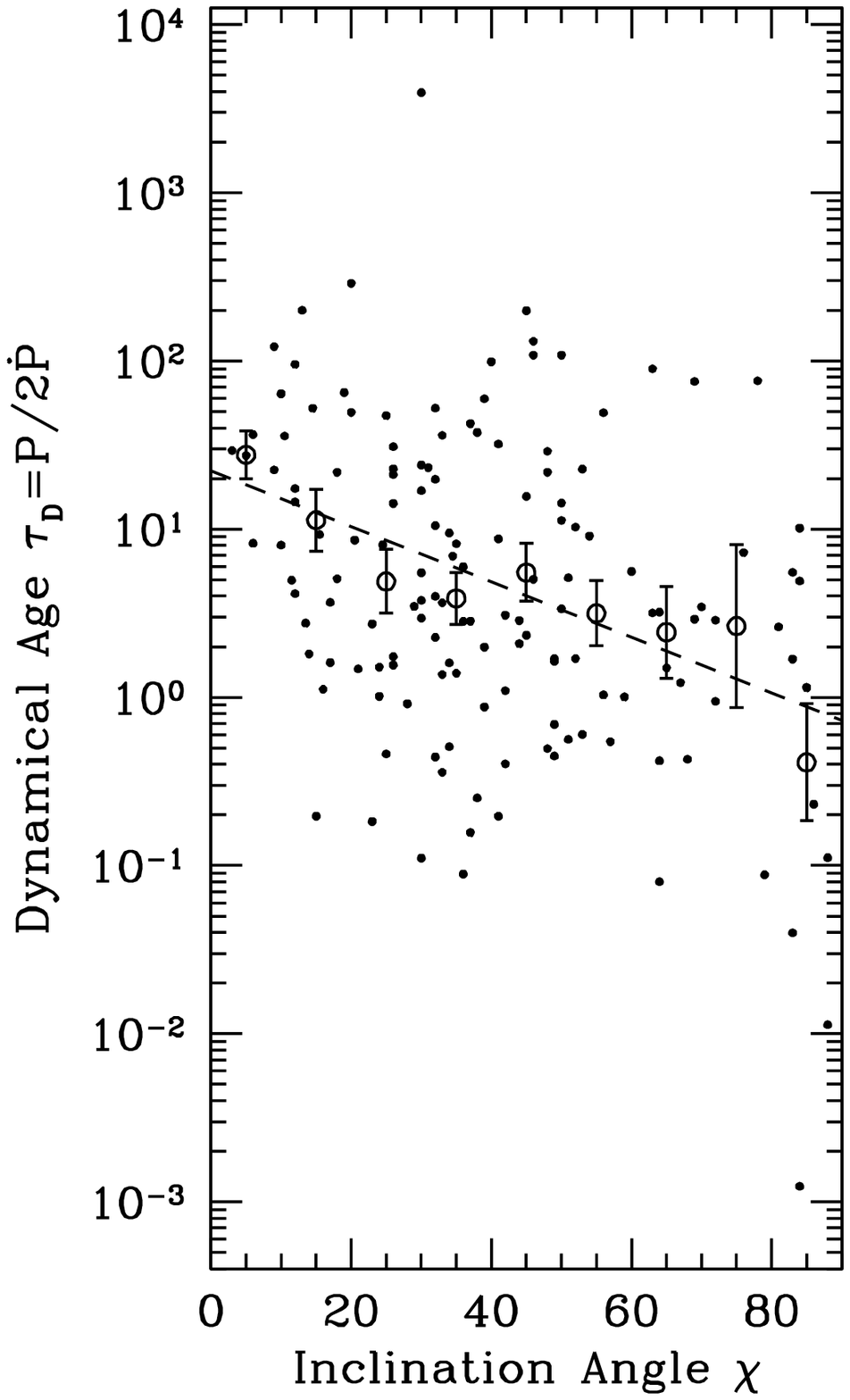}
{\caption{Dependence of the mean inclination angle $\chi$ as a function
of the pulsar dynamical age $\tau_{\rm D}$. Statistically this angle decreases with 
$P$ and, hence, with the age $\tau_{\rm D}$. After \citet{Tau98}, using data of \citet{Ran93s} and \citet{ManTable}.
        }\label{fig:05}}
\end{minipage}
\hspace{0.01\textwidth} %
\begin{minipage}[t]{0.47\linewidth}
\centering
\includegraphics[scale=0.9]{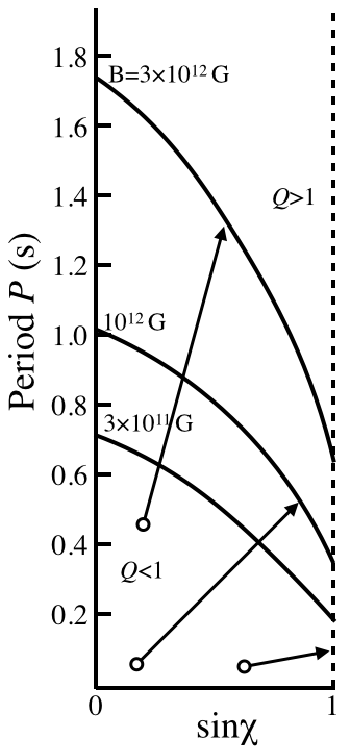}
{\caption{Pulsar extinction line in a $P-\sin\chi$ diagram for different
magnetic fields. Arrows show the evolution tracks of individual pulsars in the
model of the current losses \eqref{8bis}--\eqref{9bis}~\citep{bgi93}.
                   }\label{fig:06}}       %
\end{minipage}
\vspace{-20pt}
\end{figure}

In a careful study, \citet{Tau98} compared the inclination angles $\chi$ and rotation period $P$ for nearly 100 pulsars.
They found that $\chi$ decreases as 
$P$ increases,
over this sample of the pulsar population.
They made the straightforward assumption that the beam from the pulsar is round, as the hollow-cone model discussed in Section\ \ref{sec:HC} 
and Figure\ \ref{fig:vacuumdipole} suggest.\footnote{\citet{NarV83} suggest that pulsar beams are, instead, elongated; and that their elongation decreases as the pulsar ages.
}
They used beam radii as a function of pulse period derived by \citet{Gould} 
from comparisons among pulsars with similar periods but different impact angles $\zeta$,
and from pulsars with an interpulse (assumed to be nearly orthogonal: $\chi=\pi/2$) by \citet{rk90}.
From the observed pulse width, \citet{Tau98} then inferred the angular separation of the line of sight and the rotation axis,
and so the inclination angle $\chi$.
Figure\ \ref{fig:05} illustrates their results, using data from \citet{Ran93s} and \citet{ManTable}.
\citet{WeltevJohns} reached similar conclusions, by comparing the sample of pulsars with interpulses with the full population.

As Figure~\ref{fig:05} shows, observations reveal average statistical inclination
angles $<\chi>$ indisputably decrease as the period $P$ of pulsars
increases and its derivative $\dot P$ decreases.  Therefore, the average inclination angle
decreases as the dynamic age increases.  Correspondingly, pulsars with
longer periods exhibit relatively larger pulse widths $W_{r} = W^{(0)}_r/\sin\chi$, where $W^{(0)}_r$ is the
width of the directivity pattern~\citep{rk90,Gould,Young2010}. These results
definitely speak in favor of the alignment mechanism. On the other
hand, recently~\citet{Lyn13} on the analysis of the 45 years
observations of the Crab pulsar concluded that its inclination angle
increases with time. However, the effects of stellar
non-sphericity, leading to free precession, can account for this seemingly odd behavior \citep{Arz15b}.

The average inclination angle $\langle\chi\rangle_{\tau_{\rm D}}$ for a given range of ages $\tau_{\rm D}$ 
can decrease, even if the inclination angles of individual pulsars \emph{increases} with time, in accord with Eqn. (\ref{Intsin}).
This is a consequence of the dependence of the magnetospheric charge density on the inclination angle $\chi$.
For example, 
in the picture of \citet{rs}, radio emission results from a secondary electron-position cascade, initiated by pair-production from
curvature-radiation photons. The acceleration of primary electrons within a gap produced these curvature photons.
The Goldreich-Julian
charge density $\rho_{\rm GJ} \approx \Omega B \cos \chi/(2 \pi c)$ sets the accelerating potential across the gap.
As a pulsar ages, $\rho_{GJ}$ and the accelerating potential decline with $\Omega$.
The area of the polar cap also decreases as fewer field lines penetrate the light cylinder,
and those remaining within the polar cap are less curved.
As Eqn.\ (\ref{bh}) shows, the mean free path to pair production increases; the gap becomes wider.
When the gap width is comparable to the polar-cap radius, the cascade, and radio emission, terminate.

Because the charge density depends on $\cos \chi$ as well as $B_0$ and $\Omega$,
death comes to pulsars with different inclinations, but the same magnetic field, at different spin periods $P$.
Evaluation of the Ruderman \& Sutherland model yields $P_{max}\propto B_0^{8/9} (\cos\chi)^{2/3}$.
Indeed, as can be seen from Figure~\ref{fig:06}, for given values of the pulsar
period $P$ and the magnetic field $B_{0}$, the production of particles is suppressed
precisely at inclinations $\chi$ close to $\pi/2$, where the magnetic dipole is nearly orthogonal. 
Therefore, neutron stars 
above and to the right of the extinction lines in Figure\ \ref{fig:06} do not appear
as radio pulsars.

Because of this dependence of the pulsar extinction line on $\chi$, the
average inclination angles of the observed populations can decrease as the dynamic age increases. 
A detailed
analysis, carried out in~\citet{bgi84} (see also \citet{bn2004,be2005}) on the basis of a kinetic equation
describing the distribution of pulsars confirms this picture quantitatively.

Clearly, in any case, that the inclination angle $\chi$ is a key hidden parameter:
without taking it into account, it is impossible to construct a consistent theory of
the evolution of radio pulsars. \citet{epb2006} include this possibility in their work,
suggesting a possible direction for further improvements in models for the evolution of neutron stars~\citep{lpp1996, sgh2007, pp2007, gmvp2014}.

\section{Summary} %

Pulsars provide elegant, although not simple, laboratories for fundamental electromagnetic processes at high energies.
The basic picture of their structures, involving strong magnetic fields and rapid rotation, generation of electron-positron pairs and an energetic
wind that carries away the rotational kinetic energy of the pulsar, became clear not long after their discovery a half-century ago.
Recent work has begun to uncover the detailed structures of their magnetospheres;
the location, size, and optics of their radio emission regions;
and evolution of their spins.
Some important questions not far from solution include the effect of plasma on the magnetosphere and the possible existence of multiple states;
conversion of Poynting flux to a particle wind (the $\sigma$-problem);
the location, size, and properties of emission of different pulse components;
and whether rotation and magnetic axes tend to co-align or mis-align with time.
Further insightful theory, careful statistical studies and targeted observations will lead to deeper understanding, as pulsars continue their role as the archetypal observable neutron stars.

\begin{acknowledgements}
The authors wish to thank the International Space Science Institute for its hospitality.
V.B. and S.Ch. were supported by Russian Foundation for Basic Research 
(project N 14-02-00831) 
C.R.G. acknowledges support of the US National Science Foundation (AST-1008865). 
A.T. was supported by NASA through Einstein Postdoctoral Fellowship grant 
number PF3-140131
awarded by the Chandra X-ray Center, which is operated by the 
Smithsonian Astrophysical Observatory for NASA under contract 
NAS8-03060, and NASA via High-End Computing (HEC) Program through the 
NASA Advanced Supercomputing (NAS) Division at Ames Research Center 
that provided access to the Pleiades supercomputer, as well as NSF 
through an XSEDE computational time allocation TG-AST100040 on NICS 
Kraken, Nautilus, TACC Stampede, Maverick, and Ranch. 
\end{acknowledgements}
\bibliographystyle{aps-nameyear}      %
\bibliography{example}                %

\begin{thebibliography}{}
%
%
%
%
%
%

\bibitem[\protect\citeauthoryear{Al'ber et al.}{1975}]{ei} Al'ber, Ya.~I., Krotova, Z.~N. \& Eidman, V.~Ya.\ 1975, Astrophysics, 11, 189 %
\bibitem[\protect\citeauthoryear{Allafort et al.}{2013}]{All13} Allafort, A., Baldini, L., Ballet, J., et al.\ 2013, \apjl, 777, L2
\bibitem[\protect\citeauthoryear{Andrianov \& Beskin}{2010}]{And10} Andrianov, A.~S., Beskin, V.~S.\ 2010, Astronomy Letters, 36, 248
\bibitem[\protect\citeauthoryear{Arons \& Barnard}{1986}]{Aro86} Arons, J., Barnard, J.~J.\ 1986, \apj, 302, 120
\bibitem[\protect\citeauthoryear{Arons \& Scharlemann}{1979}]{arons3} Arons, J., Scharlemann, E.~T.\ 1979, \apj, 231, 854 %
\bibitem[\protect\citeauthoryear{Arzamasskiy et al.}{2015a}]{Arz15a} Arzamasskiy, L., Beskin, V. \& Prokofev, V.\ 2015a, \mnras, submitted, arXiv:1505.03864
\bibitem[\protect\citeauthoryear{Arzamasskiy et al.}{2015b}]{Arz15b} Arzamasskiy, L., Philippov, A. \& Tchekhovskoy, A.\ 2015b, \mnras, submitted, arXiv:1504.06626
\bibitem[\protect\citeauthoryear{Baade \& Zwicky}{1934}]{BaZwi} Baade, W., Zwicky, F.\ 1934, Proc. Nat. Acad. Sci., 20, 254 %
\bibitem[\protect\citeauthoryear{Backer}{1975}]{Bac75} Backer, D.~C.\ 1975, \aap, 43, 395
\bibitem[\protect\citeauthoryear{Backer et al.}{1976}]{brc76} Backer, D.~C., Rankin, J.~M. \& Campbell, D.~B.\ 1976, Nature, 263, 202
\bibitem[\protect\citeauthoryear{Backer \& Rankin}{1980}]{br80} Backer, D.~C., Rankin, J.~M.\ 1980, ApJS, 42, 143
\bibitem[\protect\citeauthoryear{Bailyn \& Grindlay}{1990}]{Bai90} Bailyn, C.~D., Grindlay, J.~E.\ 1990, \apj, 353, 159
\bibitem[\protect\citeauthoryear{Barnard}{1986}]{Barnard} Barnard, J.~J.,\ 1986, \apj, 303, 280
\bibitem[\protect\citeauthoryear{Barnard \& Arons}{1986}]{Bar86} Barnard, J.~J., Arons, J.\ 1986, \apj, 302, 138
\bibitem[\protect\citeauthoryear{{Belyaev}}{{Belyaev}}{2014}]{2014arXiv1412.2819B} Belyaev, M.~A.\ 2014, ArXiv:1412.2819
\bibitem[\protect\citeauthoryear{Berestetsky et al.}{1982}]{blp} Berestetsky, V.~B., Lifshits, E.~M. \& Pitaevsky L.~P. \ 1982, Relativistic Quantum Theory. Pergamon, Oxford
\bibitem[\protect\citeauthoryear{Beskin}{1999}]{be1999} Beskin, V.~S. \ 1999, Sov. Phys. Uspekhi, 41, 1071
\bibitem[\protect\citeauthoryear{Beskin \& Eliseeva}{2005}]{be2005} Beskin, V.~S., Eliseeva, S.~A. \ 2005, Astron. Lett., 31, 263
\bibitem[\protect\citeauthoryear{Beskin et al.}{1983}]{bgi83} Beskin, V.~S., Gurevich, A.~V. \& Istomin, Ya.~N. \ 1983, {\it Sov. Phys. JETP}, 58, 235 %
\bibitem[\protect\citeauthoryear{Beskin et al.}{1984}]{bgi84} Beskin, V.~S., Gurevich, A.~V. \& Istomin, Ya.~N. \ 1984, Astrophys. Space Sci., 102, 301
\bibitem[\protect\citeauthoryear{Beskin et al.}{1993}]{bgi93} Beskin, V.~S., Gurevich, A.~V. \& Istomin, Ya.~N.\ 1993, Physics of the Pulsar Magnetosphere. Cambridge University Press, Cambridge
\bibitem[\protect\citeauthoryear{Beskin et al.}{2013}]{Bes13b} Beskin, V.~S., Istomin, Y.~N. \& Philippov, A.~A.\ 2013, Physics Uspekhi, 56, 164 %
\bibitem[\protect\citeauthoryear{Beskin \& Nokhrina}{2004}]{bn2004} Beskin, V.~S., Nokhrina, E.~E. \ 2004, Astron. Lett., 30, 685
\bibitem[\protect\citeauthoryear{Beskin \& Nokhrina}{2007}]{bn2007} Beskin, V.~S., Nokhrina, E.~E. \ 2007, Astrophys. Space Sci., 308, 569
\bibitem[\protect\citeauthoryear{Beskin \& Philippov}{2011}]{BesPhil} Beskin, V.~S., Philippov, A.~A.\ 2011, ArXiv:1101.5733
\bibitem[\protect\citeauthoryear{Beskin \& Philippov}{2012}]{BesPhil2012} Beskin, V.~S., Philippov, A.~A.\ 2012, \mnras, 425, 814
\bibitem[\protect\citeauthoryear{Beskin \& Zheltoukhov}{2014}]{Bes13} Beskin, V.~S., Zheltoukhov, A.~A.\ 2014, Physics Uspekhi, 57, 799
\bibitem[\protect\citeauthoryear{Blaskiewicz  et al.}{1991}] {BCW} Blaskiewicz, M., Cordes, J.~M. \& Wasserman, I.\ 1991, Astrophys. J., 370,  643
\bibitem[\protect\citeauthoryear{Bogovalov}{1999}]{bg99} Bogovalov, S.~V.\ 1999, Astron. Astrophys., 349, 1017%
\bibitem[\protect\citeauthoryear{Bogovalov \& Khangoulyan}{2002}]{bgh1} Bogovalov, S.~V., Khangoulyan, D.~V.\ 2002, Astron. Lett, 28, 373 %
\bibitem[\protect\citeauthoryear{Boynton et al.}{1972}]{Boy72} Boynton, P.~E., Groth, E.~J., Hutchinson, D.~P., et al.\ 1972, \apj, 175, 217
\bibitem[\protect\citeauthoryear{Breton et al.}{2013}]{Bre13} Breton, R.~P., van Kerkwijk, M.~H., Roberts, M.~S.~E., et al.\ 2013, \apj, 769, 108
\bibitem[\protect\citeauthoryear{Brisken et al.}{2010}]{Bri10} Brisken, W.~F., Macquart, J.-P., Gao, J.~J., et al.\ 2010, \apj, 708, 232
\bibitem[\protect\citeauthoryear{Camenzind}{2007}]{Cmz} Camenzind, M.\ 2007, Compact Objects in Astrophysics: White Dwarfs, Neutron Stars and Black Holes. Springer, Berlin

%
%

\bibitem[\protect\citeauthoryear{Camilo et al.}{2012}]{Cam12} Camilo, F., Ransom, S.~M., Chatterjee, S., et al.\ 2012, \apj, 746, 63
\bibitem[\protect\citeauthoryear{Cerutti et al.}{2014}]{2014arXiv1410.3757C} Cerutti, B., Philippov, A., Parfrey, K. \& Spitkovsky, A.\ 2014,  ArXiv:1410.3757
\bibitem[\protect\citeauthoryear{Chen \& Beloborodov}{2013}]{Beloborodov} Chen, A.~Y., Beloborodov, A.~M. \ 2013, \apj, 762, 9 %
\bibitem[\protect\citeauthoryear{Chen \& Beloborodov}{2014}]{ChenPIC} Chen, A.~Y., Beloborodov, A.~M.\ 2014, \apjl, 795, L22
\bibitem[\protect\citeauthoryear{Cheng \& Ruderman}{1979}]{ChengRud} Cheng, A.~F., Ruderman, M.~A.\ 1979, \apj, 229, 348 %
\bibitem[\protect\citeauthoryear{Chung \& Melatos}{2011a}]{Chu11} Chung, C.~T.~Y., Melatos, A.\ 2011a, \mnras, 411, 2471 %
\bibitem[\protect\citeauthoryear{Chung \& Melatos}{2011b}]{Chu11b} Chung, C.~T.~Y., Melatos, A.\ 2011b, \mnras, 415, 1703 %
\bibitem[\protect\citeauthoryear{Cognard et al.}{1995}]{Cog95} Cognard, I., Bourgois, G., Lestrade, J.-F., et al.\ 1995, \aap, 296, 169
%
\bibitem[\protect\citeauthoryear{Cohen et al.}{1966}] {Coh66} Cohen, M.~H., Gundermann, E.~J., Hardebeck, H.~E., et al.\ 1966, Science, 153, 745 %
\bibitem[\protect\citeauthoryear{Comella et al.}{1969}]{Com69} Comella, J.~M., Craft, H.~D., Lovelace, R.~V.~E., \& Sutton, J.~M.\ 1969, \nat, 221, 453
\bibitem[\protect\citeauthoryear{Contopoulos et al.}{1999}]{ckf99} Contopoulos, I., Kazanas, D. \& Fendt, C.\ 1999, \apj, 511, 351 %
\bibitem[\protect\citeauthoryear{Cordes et al.}{1978}]{crb78} Cordes, J.~M., Rankin, J.~M. \& Backer, D.~C.\ 1978, \apj, 223, 961
\bibitem[\protect\citeauthoryear{Cordes et al.}{1983}]{cor83} Cordes, J.~M., Boriakoff, V. \& Weisberg, J.~M.\ 1983, \apj, 268, 370 %

\bibitem[\protect\citeauthoryear{Demorest et al.}{2004}]{Demo04} Demorest, P., Ramachandran, R., Backer, D.~C., et al.\ 2004, \apjl, 615, L137
\bibitem[\protect\citeauthoryear{Dodson et al.}{2007}]{Dod07} Dodson, R., Lewis, D. \& McCulloch, P.\ 2007, Astroph. Space Sci., 308, 585
\bibitem[\protect\citeauthoryear{Dyks et al.}{2004}]{Dyks04} Dyks, J., Rudak, B. \& Harding, A.~K.\ 2004, \apj, 607, 939

\bibitem[\protect\citeauthoryear{Edwards \& Stappers}{2004}] {edwsta04} Edwards, R.~T., Stappers, B.~W.\ 2004, \aap, 421, 681
\bibitem[\protect\citeauthoryear{Eliseeva et al.}{2006}]{epb2006} Eliseeva S.~A., Popov, S.~B. \& Beskin, V.~S., ArXix:astro-ph/0611320

\bibitem[\protect\citeauthoryear{Fawley et al.}{1977}]{arons1}  Fawley, W.~M., Arons, J. \& Scharlemann, E.~T.\ 1977, \apj, 217, 227%
\bibitem[\protect\citeauthoryear{Fruchter et al.}{1988}]{Fru88} Fruchter, A.~S., Stinebring, D.~R. \& Taylor, J.~H.\ 1988, \nat, 333, 237
\bibitem[\protect\citeauthoryear{Fruchter et al.}{1990}]{Fru90} Fruchter, A.~S., Berman, G., Bower, G., et al.\ 1990, \apj, 351, 642
%
\bibitem[\protect\citeauthoryear{Ghisellini et al.}{2014}]{2014Natur.515..376G} Ghisellini, G., Tavecchio, F., Maraschi, L., et al.\ 2014, \nat, 515, 376
\bibitem[\protect\citeauthoryear{Gold}{1968}]{Gold} Gold, T.\ 1968, \nat, 218, 731 %
\bibitem[\protect\citeauthoryear{Gold}{1969}]{Gold69} Gold, T.\ 1969, \nat, 221, 25
\bibitem[\protect\citeauthoryear{Goldreich \& Julian}{1969}]{GJ} Goldreich, P., Julian, W.~H.\ 1969, \apj, 157, 869
\bibitem[\protect\citeauthoryear{Gould}{1994}]{Gould} Gould, D.~M.\ 1994 PhD Thesis, University of Manchester
%
\bibitem[\protect\citeauthoryear{Gruzinov}{2005}]{gruz1} Gruzinov, A.\ 2005, Phys. Rev. Lett., 94, 021101 %
\bibitem[\protect\citeauthoryear{Gull\'{o}n et al.}{2014}]{gmvp2014} Gull\'{o}n, M., Miralles, J., Vigan\`{o}, D. \& Pons, J.\ 2014, \mnras, 443, 1891
\bibitem[\protect\citeauthoryear{Gupta et al.}{1999}]{gup99} Gupta, Y., Bhat, N.~D.~R. \& Rao, A.~P.\ 1999, 520, 173 %
\bibitem[\protect\citeauthoryear{Gwinn et al.}{1998}]{Gwi98} Gwinn, C.~R., Britton, M.~C., Reynolds, J.~E., et al.\ 1998, \apj, 505, 928 %
\bibitem[\protect\citeauthoryear{Gwinn et al.}{2000}]{gwi00} Gwinn, C.~R., Britton, M.~C., Reynolds, J.~E., et al.\ 2000, \apj, 531, 902
\bibitem[\protect\citeauthoryear{Gwinn}{2001}]{gwi01} Gwinn, C.~R.\ 2001, \apj, 554, 1197 %
\bibitem[\protect\citeauthoryear{Gwinn et al.}{2011}]{Gwi11} Gwinn, C.~R., Johnson, M.~D., Smirnova, T.~V., \& Stinebring, D.~R.\ 2011, \apj, 733, 52
%
\bibitem[\protect\citeauthoryear{Gwinn et al.}{2012a}]{gwi12a} Gwinn, C.~R., Johnson, M.~D., Reynolds, J.~E.,
%
%
et al.\ 2012a, \apj, 758, 6 %
\bibitem[\protect\citeauthoryear{Gwinn et al.}{2012b}] {gwi12b} Gwinn, C.~R., Johnson, M.~D., Reynolds, J.~E.,
%
%
et al.\ 2012b, \apj, 758, 7 %

\bibitem[\protect\citeauthoryear{Hakobyan \& Beskin}{2014}]{HaBes14} Hakobyan, H.~L. \& Beskin, V.~S.\ 2014, Astronomy Reports, 58, 889
%
\bibitem[\protect\citeauthoryear{Hankins \& Rankin}{2010}]{HankinRankin} Hankins, T.~H., Rankin, J.~M.\ 2010, Astron. J., 139, 168
\bibitem[\protect\citeauthoryear{Harding et al.}{2008}]{Har08} Harding, A.~K., Stern, J.~V., Dyks, J. \& Frackowiak, M.\ 2008, \apj, 680, 1378
\bibitem[\protect\citeauthoryear{Helfand et al.}{1980}]{Helf80} Helfand, D.~J., Taylor, J.~H., Backus, P.~R., \& Cordes, J.~M.\ 1980, \apj, 237, 206
\bibitem[\protect\citeauthoryear{Hewish et al.}{1968}]{nature1968} Hewish, A., Bell, S.~J., Pilkington, J.D. et al.\ 1968, Nature, 217, 709 %
\bibitem[\protect\citeauthoryear{Hewish et al.}{1974}] {Hew74} Hewish, A., Readhead, A.~C.~S. \& Duffett-Smith, P.~J.\ 1974, \nat, 252, 657 %
\bibitem[\protect\citeauthoryear{Hirano \& Gwinn}{2001}]{hir01} Hirano, C., Gwinn, C.~R.\ 2001, \apj, 553, 358 %
\bibitem[\protect\citeauthoryear{Ingraham}{1973}]{ingr} Ingraham, R.\ 1973, \apj, 186, 625 %
%
\bibitem[\protect\citeauthoryear{Istomin \& Sobyanin}{2011a}]{Sobyanin1} Istomin, Y.~N., Sobyanin, D.~N.\ 2011, JETP, 113, 592 %
\bibitem[\protect\citeauthoryear{Istomin \& Sobyanin}{2011b}]{Sobyanin2} Istomin, Y.~N., Sobyanin, D.~N.\ 2011, JETP, 113, 605 %
\bibitem[\protect\citeauthoryear{Jenet \& Ransom}{2004}]{Jen04} Jenet, F.~A., Ransom, S.~M.\ 2004, \nat, 428, 919
%
\bibitem[\protect\citeauthoryear{Johnson \& Gwinn}{2012a}] {joh12a} Johnson, M.~D., Gwinn, C.~R.\ 2012a, \apj, 755, 179 %
\bibitem[\protect\citeauthoryear{Johnson et al.}{2012b}] {joh12b} Johnson, M.~D., Gwinn, C.~R. \& Demorest, P.\ 2012b, \apj, 758, 8%
\bibitem[\protect\citeauthoryear{Johnson \& Gwinn}{2013}]{Joh13} Johnson, M.~D., Gwinn, C.~R.\ 2013, \apj, 768, 170

\bibitem[\protect\citeauthoryear{Kalapotharakos \& Contopoulos}{2009}]{contop1} Kalapotharakos, C., Contopoulos, I.\ 2009, Astron. Astrophys., 496, 495 %
\bibitem[\protect\citeauthoryear{Kalapotharakos et al.}{2012}]{kalap12} Kalapotharakos, C., Contopoulos, I., Kazanas, D.,\ 2012, \mnras, 420,  2793
\bibitem[\protect\citeauthoryear{Kalapotharakos et al.}{2012}]{contop2} Kalapotharakos, C., Kazanas, D., Harding, A. \& Contopoulos, I.\ 2012 \apj, 749, 2 %
\bibitem[\protect\citeauthoryear{Kaplan et al.}{2013}]{Kap13} Kaplan, D.~L., Bhalerao, V.~B., van Kerkwijk, M.~H., et al.\ 2013, \apj, 765, 158
\bibitem[\protect\citeauthoryear{Kardashev et al.}{2013}]{Kard13} Kardashev, N.~S., Khartov, V.~V., Abramov, V.~V., et al.\ 2013, Astronomy Reports, 57, 153
\bibitem[\protect\citeauthoryear{Kaspi et al.}{1994}]{Kas94} Kaspi, V.~M., Taylor, J.~H. \& Ryba, M.~F.\ 1994, \apj, 428, 713
%
\bibitem[\protect\citeauthoryear{Kennel \& Coroniti}{1984a}]{kc1} Kennel, C.~F., Coroniti, F.~V.\ 1984a, \apj, 283, 694%
\bibitem[\protect\citeauthoryear{Kennel \& Coroniti}{1984b}]{kc2} Kennel, C.~F., Coroniti, F.~V.\ 1984b, \apj, 283, 710%
\bibitem[\protect\citeauthoryear{Kern et al.}{2000}] {ker00} Kern, J.~S., Hankins, T.~H. \& Rankin, J.~M.\ 2000, in \textit{IAU Colloq.~177: Pulsar Astronomy - 2000 and Beyond}, eds. M. Kramer, N. Wex, and N. Wielebinski, 202, 257
\bibitem[\protect\citeauthoryear{Kirk et al.}{2009}]{Kirk09} Kirk, J.~G., Lyubarsky, Y. \& Petri, J.\ 2009, Astrophys. and Space Sci. Library, 357, 421
\bibitem[\protect\citeauthoryear{Komesaroff et al.}{1974}]{kom74} Komesaroff, M.~M., McCulloch, P.~M. \& Rankin, J.~M. \ 1974, \nat, 252, 210 %
\bibitem[\protect\citeauthoryear{Komissarov}{2006}]{k06} Komissarov, S.~S.\ 2006, \mnras, 367, 19%
\bibitem[\protect\citeauthoryear{Komissarov et al.}{2009}]{2009MNRAS.394.1182K} Komissarov, S.~S., Vlahakis, N., K{\"o}nigl, A., \& Barkov, M.~V.\ 2009, \mnras, 394, 1182
\bibitem[\protect\citeauthoryear{Komissarov \& Lyubarsky}{2003}]{kl03}  Komissarov, S.~S., Lyubarsky, Yu.~E.\ 2003, \mnras, 344, L93 %
%
\bibitem[\protect\citeauthoryear{Kramer et al.}{2006}]{Kra06} Kramer, M., Lyne, A.~G., O'Brien, J.~T., et al.\ 2006, Science, 312, 549 %
\bibitem[\protect\citeauthoryear{Kravtsov \& Orlov}{1980}]{KravtsovOrlov} Kravtsov, Yu.~A., Orlov, Yu.~I.,\ 1980, Geometricheskaya optika neodnorodnykh sred (Geometrical Optics of Inhomogeneous Media) (Moscow, Nauka, 1980) [Translated inti English (Berlin: Springer-Verlag, 1990)]%
\bibitem[\protect\citeauthoryear{Krzeszowski et al.}{2009}]{Krzes} Krzeszowski, K., Mitra, D., Gupta, Y., et al.\ 2009, \mnras, 393, 1617
\bibitem[\protect\citeauthoryear{Kulkarni et al.}{1988}]{Kulk88} Kulkarni, S.~R., Djorgovski, S. \& Fruchter, A.~S.\ 1988, \nat, 334, 504
\bibitem[\protect\citeauthoryear{Landau}{1932}]{Landau} Landau, L.~D.\ 1932, Phys. Zeit. Sow., 1, 271 %
\bibitem[\protect\citeauthoryear{Landau \& Lifshitz}{1989}]{LLField} Landau, L.~D., Lifshitz, E.~M.\ 1989, The Classical Theory of Fields, Pergamon Press, Oxford
\bibitem[\protect\citeauthoryear{Li et al.}{2012}]{Li12} Li, J., Spitkovsky, A. \& Tchekhovskoy, A.\ 2012a, \apj, 746, 60 %
\bibitem[\protect\citeauthoryear{Li et al.}{2012}]{Li12b} Li, J., Spitkovsky, A. \& Tchekhovskoy, A.\ 2012b, \apjl, 746, L24 %
\bibitem[\protect\citeauthoryear{Lipunov et al.}{1996}]{lpp1996} Lipunov, V.~M., Postnov, K.~A. \& Prokhorov, M.~E. \ 1996 Astron. Astophys., 310, 489
\bibitem[\protect\citeauthoryear{Lommen et al.}{2007}]{Lomm07} Lommen, A., Donovan, J., Gwinn, C., et al.\ 2007, \apj, 657, 436
%
\bibitem[\protect\citeauthoryear{Lyne \& Manchester}{1988}]{Lyn88} Lyne, A.~G., Manchester, R.~N.\ 1988, \mnras, 234, 477
\bibitem[\protect\citeauthoryear{Lyne \& Graham-Smith}{1998}]{lg-s} Lyne, A.~G., Graham-Smith, F.\ 1998, Pulsar Astronomy, Cambridge University Press, Cambridge
\bibitem[\protect\citeauthoryear{Lyne et al.}{2004}]{Lyn04} Lyne, A.~G., Burgay, M., Kramer, M., et al.\ 2004, Science, 303, 1153
\bibitem[\protect\citeauthoryear{Lyne}{2009}]{Lyn09} Lyne, A.~G.\ 2009, Astrophys. and Space Sci. Library, 357, 67
\bibitem[\protect\citeauthoryear{Lyne et al.}{2010}]{Lyn10} Lyne, A., Hobbs, G., Kramer, M., et al.\ 2010, Science, 329, 408 %
\bibitem[\protect\citeauthoryear{Lyne et al.}{2013}]{Lyn13} Lyne, A., Graham-Smith, F., Weltevrede, P., et al.\ 2013, Science, 342, 598 %
\bibitem[\protect\citeauthoryear{Lyubarsky}{2010}]{2010MNRAS.402..353L} Lyubarsky, Y.~E.\ 2010, \mnras, 402, 353
\bibitem[\protect\citeauthoryear{Lyutikov et al.}{1999}]{Lyu99} Lyutikov, M., Blandford, R.~D. \& Machabeli, G.\ 1999, \mnras, 305, 338
\bibitem[\protect\citeauthoryear{Lyutikov \& Parikh}{2000}] {lyu00} Lyutikov, M., Parikh, A.\ 2000, ApJ, 541, 1016 %
\bibitem[\protect\citeauthoryear{Lyutikov}{2004}]{Lyu04} Lyutikov, M.\ 2004, \mnras, 353, 1095

\bibitem[\protect\citeauthoryear{Manchester}{1995}]{Man95} Manchester, R.~N.\ 1995, Journal of Astrophysics and Astronomy, 16, 107
\bibitem[\protect\citeauthoryear{Manchester et al.}{1975}]{mth75} Manchester, R.~N., Taylor, J.~H. \& Huguenin, G.~R.\ 1975, \apj, 196, 83
\bibitem[\protect\citeauthoryear{Manchester \& Taylor}{1977}]{ManTay} Manchester, R.~N., Taylor, J.~H.\ 1977, Pulsars (San Francisco:W.H. Freeman)
\bibitem[\protect\citeauthoryear{Manchester et al.}{2005}]{ManTable} Manchester, R.~N., Hobbs, G.~B., Teoh, A. \& Hobbs, M.,\ 2005, AJ, 129, 1993
\bibitem[\protect\citeauthoryear{McKinney}{2006}]{mck06} McKinney, J.C.\ 2006, \mnras, 368, L30%
\bibitem[\protect\citeauthoryear{McKinnon}{2003}]{mck03a} McKinnon, M.~M.\ 2003, \apj, 590, 1026
\bibitem[\protect\citeauthoryear{McKinnon}{2009}]{McK09} McKinnon, M.~M.\ 2009, \apj, 692, 459
\bibitem[\protect\citeauthoryear{McLaughlin et al.}{2004}]{McLa04} McLaughlin, M.~A., Kramer, M., Lyne, A.~G., et al.\ 2004, \apjl, 613, L57  %
\bibitem[\protect\citeauthoryear{Medin \& Lai}{2010}]{Laietalcap} Medin, Z., Lai, D.\ 2010, \mnras, 406, 1379 %
\bibitem[\protect\citeauthoryear{Mestel et al.}{1999}]{mps99} Mestel, L., Panagi, P. \& Shibata, S.\ 1999, \mnras, 309, 388 %
\bibitem[\protect\citeauthoryear{Michel}{1973}]{mich73b} Michel, F.~C.\ 1973, \apj, 180, L133 %
\bibitem[\protect\citeauthoryear{Michel}{1974}]{mich74} Michel, F.~C.\ 1974, \apj, 187, 585%
\bibitem[\protect\citeauthoryear{Mikhailovskii et al.}{1982}]{Mikh82} Mikhailovskii A.~B. et al.\ 1982, Sov. Astron. Lett., 8, 369
\bibitem[\protect\citeauthoryear{Mitra \& Li}{2004}] {ML2004} Mitra, D., Li, X.~H.\ 2004, Astron. Astrophys., 421, 15-228
\bibitem[\protect\citeauthoryear{Narayan \& Vivekanand}{1983}]{NarV83} Narayan, R., Vivekanand, M.\ 1983, \aap, 122, 45
\bibitem[\protect\citeauthoryear{Nomoto \& Kondo}{1991}]{Nom91} Nomoto, K., Kondo, Y.\ 1991, \apjl, 367, L19
\bibitem[\protect\citeauthoryear{Ogura \& Kojima}{2003}]{ok03} Ogura, J., Kojima, Y.\ 2003, Prog. Theor. Phys., 109, 619%
\bibitem[\protect\citeauthoryear{Pacini}{1967}]{Pacini} Pacini, F.\ 1967, \nature, 216, 567%
\bibitem[\protect\citeauthoryear{Pallanca et al.}{2012}]{Pall12} Pallanca, C., Mignani, R.~P., Dalessandro, E., et al.\ 2012, \apj, 755, 180
\bibitem[\protect\citeauthoryear{Parfrey et al.}{2012}]{par12phaedra} Parfrey, K., Beloborodov, A.~M. \& Hui, L.,\ 2012, \mnras, 423, 1416
\bibitem[\protect\citeauthoryear{Pen et al.}{2014}]{Pen14} Pen, U.-L., Macquart, J.-P., Deller, A.~T., \& Brisken, W.\ 2014, \mnras, 440, L36
\bibitem[\protect\citeauthoryear{Perera et al.}{2015}]{Per15} Perera, B.~B.~P., Stappers, B.~W., Weltevrede, P. et al.\ 2015, \mnras, 446, 1380
\bibitem[\protect\citeauthoryear{P{\'e}tri}{2012}]{petri12a} P{\'e}tri, J.,\ 2012, \mnras, 424, 605
\bibitem[\protect\citeauthoryear{Petrova}{2006}]{Petrova} Petrova, S.,\ 2006, \mnras, 368, 1764%
\bibitem[\protect\citeauthoryear{Petrova \& Lyubarskii}{2000}]{PetrovaLyu} Petrova, S., Lyubarskii, Yu.\ 2000, Astron. Astrophys., 355, 1168
\bibitem[\protect\citeauthoryear{Philippov et al.}{2014}]{Phi13} Philippov, A., Tchekhovskoy, A. \& Li, J.~G.\ 2014, \mnras, 441, 1879  %
\bibitem[\protect\citeauthoryear{Philippov et al.}{2014}]{2014arXiv1412.0673P} Philippov, A.~A., Spitkovsky, A. \& Cerutti, B.,\ 2014b, ArXiv:1412.0673
\bibitem[\protect\citeauthoryear{Philippov \& Spitkovsky}{2014}]{2014ApJ...785L..33P} Philippov, A.~A., Spitkovsky, A.,\ 2014, \apjl, 785, L33
\bibitem[\protect\citeauthoryear{Popov et al.}{2006}]{pop06a} Popov, M.~V., Soglasnov, V.~A., Kondratiev, V.~I., et al.\ 2006, Astronomy Reports, 50, 55 %
%
\bibitem[\protect\citeauthoryear{Popov \& Prokhorov}{2007}]{pp2007} Popov, S.~B., Prokhorov, M.~E. \ 2007,  Phys. Usp., 50, 1123
\bibitem[\protect\citeauthoryear{Radhakrishnan et al.}{1969}]{Radetal69} Radhakrishnan, V., Cooke, D.~J., Komesaroff, M.~M., \& Morris, D.\ 1969, \nat, 221, 443
\bibitem[\protect\citeauthoryear{Radhakrishnan \& Cooke}{1969}]{RadCoo69} Radhakrishnan, V., Cooke, D.~J.\ 1969, Astrophys. Lett., 3, 225 %
\bibitem[\protect\citeauthoryear{Rankin}{1983}]{rk83} Rankin, J.~M.\ 1983, \apj, 274, 333%
\bibitem[\protect\citeauthoryear{Rankin}{1986}]{Ran86} Rankin, J.~M.\ 1986, \apj, 301, 901
\bibitem[\protect\citeauthoryear{Rankin}{1990}]{rk90} Rankin, J.~M.\ 1990, \apj, 352, 247 %
\bibitem[\protect\citeauthoryear{Rankin}{1993}]{Ran93s} Rankin, J.~M.\ 1993, \apjs, 85, 145
\bibitem[\protect\citeauthoryear{Rankin}{1993}]{ran93} Rankin, J.~M.\ 1993, \apj, 405, 285 %
\bibitem[\protect\citeauthoryear{Readhead \& Hewish}{1972}] {Rea72} Readhead, A.~C.~S., Hewish, A.\ 1972, \nat, 236, 440 %
\bibitem[\protect\citeauthoryear{Richards \& Comella}{1969}]{RichCom69} Richards, D.~W., Comella, J.~M.\ 1969, \nat, 222, 551 %
\bibitem[\protect\citeauthoryear{Roberts}{2011}]{Rob11} Roberts, M.~S.~E.\ 2011, in \textit{American Institute of Physics Conference Series}, 1357, 127
\bibitem[\protect\citeauthoryear{Romani et al.}{2012}]{Rom12} Romani, R.~W., Filippenko, A.~V., Silverman, J.~M., et al.\ 2012, \apjl, 760, LL36
\bibitem[\protect\citeauthoryear{Ruderman \& Sutherland}{1975}]{rs} Ruderman, M.~A., Sutherland, P.~G.\ 1975, \apj, 196, 51 %
\bibitem[\protect\citeauthoryear{Ruiz et al.}{2014}]{2014PhRvD..89h4045R} Ruiz, M., Paschalidis, V. \& Shapiro, S.~L.,\ 2014, \prd, 89, 084045
\bibitem[\protect\citeauthoryear{Ryba \& Taylor}{1991}]{RyTa91} Ryba, M.~F., Taylor, J.~H.\ 1991, \apj, 380, 557
\bibitem[\protect\citeauthoryear{Scharlemann et al.}{1978}]{arons2} Scharlemann, E.~T., Fawley, W.~M. \& Arons, J.\ 1978, \apj, 222, 297 %
\bibitem[\protect\citeauthoryear{Scheuer}{1968}] {sch68} Scheuer, P.~A.~G.\ 1968, \nat, 218, 920
\bibitem[\protect\citeauthoryear{Schwab et al.}{2015}]{Schwab} Schwab, J., Quataert, E. \& Bildsten, L.\ 2015, \mnras, in press %
\bibitem[\protect\citeauthoryear{Scott et al.}{2003}]{Scott03} Scott, D.~M., Finger, M.~H. \& Wilson, C.~A.\ 2003, \mnras, 344, 412
\bibitem[\protect\citeauthoryear{Shapiro \& Teukolsky}{1985}]{ShapTeuk} Shapiro, S.~L., Teukolsky, S.~A.\ 1985, Black Holes, White Dwarfs, and Neutron Stars. A Wiley--Interscience Publication, New York
\bibitem[\protect\citeauthoryear{Shearer et al.}{2003}]{Shea03} Shearer, A., Stappers, B., O'Connor, P., et al.\ 2003, Science, 301, 493
\bibitem[\protect\citeauthoryear{Smirnova et al.}{1996}] {smi96} Smirnova, T.~V., Shishov, V.~I. \& Malofeev, V.~M.\ 1996, \apj, 462, 289 %
\bibitem[\protect\citeauthoryear{Spitkovsky}{2006}]{Spi06} Spitkovsky, A.\ 2006, \apj, 648, L51%
\bibitem[\protect\citeauthoryear{Stappers et al.}{1996}]{Stapp96} Stappers, B.~W., Bessell, M.~S. \& Bailes, M.\ 1996, \apjl, 473, L119
\bibitem[\protect\citeauthoryear{Stinebring et al.}{1984a}]{Sti84a} Stinebring, D.~R., Cordes, J.~M., Rankin, J.~M., et al.\ 1984, \apjs, 55, 247 %
\bibitem[\protect\citeauthoryear{Stinebring et al.}{1984b}]{Sti84b} Stinebring, D.~R., Cordes, J.~M., Weisberg, J.~M., et al.\ 1984, \apjs, 55, 279
\bibitem[\protect\citeauthoryear{Story et al.}{2007}]{sgh2007} Story, S.~A., Gonthier, P.~L. \& Harding, A. \ 2007, \apj, 671, 713
\bibitem[\protect\citeauthoryear{Strader et al.}{2013}]{Stra13} Strader, M.~J., Johnson, M.~D., Mazin, B.~A., et al.\ 2013, \apjl, 779, LL12
\bibitem[\protect\citeauthoryear{Sturrock}{1971}]{stur71} Sturrock, P.~A.\ 1971, \apj, 164, 529 %

\bibitem[\protect\citeauthoryear{Tauris \& Manchester}{1998}]{Tau98} Tauris, T.~M., Manchester, R.~N.\ 1998, \mnras, 298, 625 %
                                %
\bibitem[\protect\citeauthoryear{Tchekhovskoy et al.}{2009}]{Tch09} Tchekhovskoy, A., McKinney, J.~C. \& Narayan, R.\ 2009, \apj, 699, 1789 %
\bibitem[\protect\citeauthoryear{Tchekhovskoy et al.}{2010}]{2010NewA...15..749T} Tchekhovskoy, A., Narayan, R. \& McKinney, J.~C.\ 2010, New Astronomy, 15, 749
\bibitem[\protect\citeauthoryear{Tchekhovskoy et al.}{2011}]{2011MNRAS.418L..79T} Tchekhovskoy, A., Narayan, R. \& McKinney, J.~C.\ 2011, \mnras, 418, L79
\bibitem[\protect\citeauthoryear{Tchekhovskoy \& McKinney}{2012}]{2012MNRAS.423L..55T} Tchekhovskoy, A., McKinney, J.~C.\ 2012, \mnras, 423, L55
\bibitem[\protect\citeauthoryear{Tchekhovskoy et al.}{2013}]{Tch13} Tchekhovskoy, A., Spitkovsky, A., \& Li, J.~G.\ 2013, \mnras, 435, L1 %
\bibitem[\protect\citeauthoryear{Tchekhovskoy et al.}{2015}]{Tch15} Tchekhovskoy, A., Philippov, A. \& Spitkovsky, A.\ 2015, \mnras,
  submitted, arXiv:1503.01467
\bibitem[\protect\citeauthoryear{Tchekhovskoy}{2015}]{2015ASSL..414...45T}
  Tchekhovskoy, A.\ 2015, Astrophys. and Space Sci. Library, 414, 45, doi:10.1007/978-3-319-10356-3\_3
\bibitem[\protect\citeauthoryear{Thompson et al.}{2001}]{tms01} Thompson, A.~R., Moran, J.~M. \& Swenson, G.~W., Jr.\ Interferometry and synthesis in radio astronomy~2nd ed.~ (New York : Wiley 2001)%
\bibitem[\protect\citeauthoryear{Timokhin}{2006}]{tim06} Timokhin, A.~N.\ 2006, \mnras, 368, 1055%
\bibitem[\protect\citeauthoryear{Timokhin}{2010}]{Tim} Timokhin, A.~N.\ 2010, \mnras, 408, 2092 %
\bibitem[\protect\citeauthoryear{Timokhin \& Arons}{2013}]{TimArons} Timokhin, A.~N., Arons, J.\ 2013, \mnras, 429, 20 %
\bibitem[\protect\citeauthoryear{Walker et al.}{2008}]{Wal08} Walker, M.~A., Koopmans, L.~V.~E., Stinebring, D.~R., \& van Straten, W.\ 2008, \mnras, 388, 1214
\bibitem[\protect\citeauthoryear{Wang et al.}{2010}]{Wang} Wang, C., Lai, D., \& Han, J.,\ 2010, \mnras, 403, 569 %
\bibitem[\protect\citeauthoryear{Weltevrede \& Johnston}{2008}]{WeltevJohns} Weltevrede, P., Johnston, S.\ 2008, \mnras, 391, 1210

\bibitem[\protect\citeauthoryear{Whelan \& Iben}{1973}]{Whe73} Whelan, J., Iben, I.\ 1973, \apj, 186, 1007
\bibitem[\protect\citeauthoryear{Wolszczan \& Cordes}{1987}]{wol87} Wolszczan, A., Cordes, J.~M.\ 1987, ApJ 320, L35 %
\bibitem[\protect\citeauthoryear{Young et al.}{2010}]{Young2010} Young, M.~D.~T., et al. \ 2010, \mnras, 402, 1317
\bibitem[\protect\citeauthoryear{Yuen \& Melrose}{2014}]{YuMe} Yuen, R., Melrose, D.~B.\ 2014, Publications of the Astronomical Society of Australia, 31, 39
\bibitem[\protect\citeauthoryear{Zamaninasab et al.}{2014}]{2014Natur.510..126Z} Zamaninasab, M., Clausen-Brown, E., Savolainen, T.
\& Tchekhovskoy, A.\ 2014, \nat, 510, 126
\bibitem[\protect\citeauthoryear{Zdziarski et al.}{2014}]{2014arXiv1410.7310Z} Zdziarski, A.~A., Sikora, M., Pjanka, P. \& Tchekhovskoy, A.\ 2014, arXiv:1410.7310




%
%


%
%
%
%
%
\end{thebibliography}
\nocite{*}

\clearpage

\end{document}